\documentclass[twocolumn,traditabstract]{aa}

\usepackage{txfonts}
\usepackage{graphicx}
\usepackage{natbib}
\usepackage{hyperref}

\renewcommand{\deg}{\mbox{$^{\circ}$}}

\begin{document}

   \title{The long-wavelength thermal emission of the Pluto-Charon system from {\em Herschel} \thanks{{\em Herschel} is an ESA space observatory with science instruments provided
by European-led Principal Investigator consortia and with important participation from NASA.}  observations.   Evidence for emissivity effects.}

%   \subtitle{I. Overviewing the $\kappa$-mechanism}

\author{E. Lellouch\inst{1}
\and P.~Santos-Sanz\inst{2}
\and S.~Fornasier\inst{1}
\and T.~Lim\inst{3}
\and J.~Stansberry\inst{4}
\and E.~Vilenius\inst{5,6}
\and Cs. Kiss\inst{7}
\and T.~M\"uller\inst{6}
\and G.~Marton\inst{6}
\and S.~Protopapa\inst{8}
\and P.~Panuzzo\inst{9}
\and R.~Moreno\inst{1}
}

 \institute{LESIA, Observatoire de Paris, PSL Research University, CNRS, Sorbonne Universit\'es, UPMC Univ. Paris 06, Univ. Paris Diderot, Sorbonne Paris Cit\'e, 5 place Jules Janssen, 92195 Meudon, France\\ %1
\email{emmanuel.lellouch@obspm.fr}
\and Instituto de Astrof\'isica de Andaluc\'ia-CSIC, Glorieta de la Astronom\'ia s/n, 18008-Granada, Spain.
\and European Space Astronomy Centre (ESAC), P.O. Box 78, E-28691 Villanueva de la Ca\~nada, Madrid, Spain                                         
\and Space Telescope Science Institute, 3700 San Martin Drive, Baltimore, MD 21218 USA
\and Max-Planck-Institut f\"{u}r Sonnensystemforschung, Justus-von-Liebig-Weg 3, 37077 G\"ottingen, Germany         
\and Max-Planck-Institut f\"ur Extraterrestrische Physik, 
Giessenbachstra\ss e, 85748 Garching, Germany   
\and Konkoly Observatory of the Hungarian Academy of Sciences, H-1121 Budapest, Konkoly Thege Miklós út 15-17, Hungary
\and Department of Astronomy, University of Maryland, College Park, MD 20742, USA
\and GEPI, Observatoire de Paris, PSL Research University, CNRS,
Univ. Paris Diderot, Sorbonne Paris Cit\'e, 5 Place Jules Janssen, 92195
Meudon, France                                       %15
             }

   \titlerunning{Herschel observations of Pluto}
   \authorrunning{Lellouch et al.}

   \date{Received, November 2, 2015; Revised, December 19, 2015}

%   \date{\today}

  \abstract
   {Thermal observations of the Pluto-Charon system acquired by the {\em Herschel} Space Observatory in
February 2012 are presented. They consist of photometric measurements with the PACS and SPIRE instruments
(nine visits to the Pluto system each), covering  six wavelengths from 70 to 500 $\mu$m altogether. The thermal light curve
of Pluto-Charon is observed in all filters, albeit more marginally at 160 and especially 500 $\mu$m. Putting these data
into the context of older ISO, {\em Spitzer} and ground-based observations indicates that the brightness
temperature (T$_B$) of the system (rescaled to a common heliocentric distance) drastically decreases with increasing 
wavelength, from $\sim$53 K at 20 $\mu$m to $\sim$35 K at 500 $\mu$m, and perhaps ever less at longer
wavelengths. Considering a variety of diurnal and/or seasonal thermophysical models, we show that
T$_B$ values of 35 K are lower than any expected temperature for the dayside surface or subsurface
of Pluto and Charon, implying a low surface emissivity. Based on multiterrain modeling, we infer a spectral emissivity that decreases steadily from
1 at 20-25 $\mu$m to $\sim$0.7 at 500~$\mu$m. This kind of behavior is usually not observed in asteroids (when proper 
allowance is made for subsurface sounding), but is found in several icy surfaces of the solar system. We tentatively
identify that a combination of a strong dielectric constant and a considerable surface material transparency (typical penetration
depth $\sim$1 cm) is responsible for the effect. Our results have implications for the interpretation
of the temperature measurements by REX/{\em New Horizons} at 4.2 cm wavelength.
  
}


\maketitle

%----------------------------------------------------------------------
\section{Introduction}
The {\em New Horizons} flyby of the Pluto system on July 14, 2015 revealed Pluto and Charon as planetary worlds \citep{stern15}.
Pluto appears to display an unexpected variety of terrain morphologies, suggesting a complex history of surface activity. These include
icy plains with evidence for glacier-like flows of ice and polygonal ice patterns, mountain ridges several kilometers
high, and dark, cratered, ancient terrains, where irradiation of surface ices (N$_2$, CH$_4$, CO) and/or atmospheric production of organic tholins falling to the surface may be responsible for the dark red color. While the identification of the
processes shaping this rich geology is just beginning, it already seems clear that Pluto's surface appearance is to a large extent sculpted by interactions between its mobile volatile ices, evolving N$_2$-dominated atmosphere, and surface bedrock. Mars-like 
seasonal cycles must be at work, in which volatile N$_2$ (and the secondary species CH$_4$
and CO) are shared between atmospheric and surface ice reservoirs through sublimation/condensation exchanges and 
volatile migration. These processes are related to the temperature distribution across Pluto's surface, which reflects
the balance between insolation, thermal radiation, thermal conduction, and latent heat exchanges, and depends on important parameters such as albedo, emissivity, and thermal inertia  \citep[see, e.g.,][]{hansen96,young12}. At Charon, where no atmosphere has yet been detected, such resurfacing processes are less obvious, although the distinctly red color of Charon's north polar region may be related to seasonal cold trapping of volatiles in that region, followed by energetic radiation \citep{stern15}.

Temperature measurements on an icy surface are possible from the temperature-dependent position and shape of near-IR
absorption bands \citep[e.g.,][]{quirico97,tryka94,tryka95,grundy99,grundy02}. This diagnostic is to 
be used by {\em New Horizons}/Ralph \citep{reuters08}, for example, for the N$_2$(2-0) ice band at 2.15 $\mu$m.
The only other method for determining surface temperatures is thermal radiometry. Thermal measurements (in general
spatially unresolved) of the Pluto system at a variety of wavelengths 
(from $\sim$20 to 1400~$\mu$m) have been acquired using IRAS, ISO, and {\em Spitzer}, and a number of ground-based
mm/submm facilities. In particular, the ISO and {\em Spitzer} measurements clearly
detected the Pluto+Charon thermal light curve that is associated with the albedo contrasts on Pluto and the diurnal
variability of insolation. These measurements have provided the first determination of the thermal inertia 
of Pluto and Charon, and some constraints on their emissivity behavior over 20-160 $\mu$m \citep{lellouch00a,lellouch11}.

The operation of {\em Herschel} \citep{pilbratt10} in 2009-2013 offered an opportunity to extend the study toward longer wavelengths (70-500 $\mu$m), bridging the gap with the sub-mm/mm measurements. Combined with previous
{\em Spitzer} data, these measurements permit us to refine our estimates of Pluto's and Charon's thermal inertia,
and determine the long-wavelength behavior of the system's emission. Following an initial assessment of the data  
\citep{lellouch13a}, we  present a detailed report of these observations and their modeling. The {\em Herschel} 70-$\mu$m data that are described below have already been used to derive limits on the amount of 
dust in the Pluto--Charon system \citep{marton15}.

\begin{table*}
%\begin{minipage}[t]{\columnwidth}
\caption{Summary of observations} % title of Table
\label{obssum}      % is used to refer this table in the text
\centering                          % used for centering table
\renewcommand{\footnoterule}{} 
\footnotesize\addtolength{\tabcolsep}{-4pt}
\begin{center}
\begin{tabular}{lccccccl}
\noalign{\smallskip} \hline \noalign{\smallskip}
Obs. ID  & Instrument/Mode & Filter  & Start Time          &  T$_{obs}$& $\Delta^a$ & Longitude$^b$ & Flux$^c$\\
 & & & (2012-)&   (sec.) & & & (mJy) \\
\noalign{\smallskip} \hline \noalign{\smallskip}
1342239786  &  SPIREPhoto & 250/350/500 & 02-29 20:08:23 & 1421 & 32.649  & 28.0 & 179.9$\pm$2.7 / 107.9$\pm$2.9 /  ~58.4$\pm$3.5\\
1342239900  &  SPIREPhoto & 250/350/500 & 03-01 13:04:38 & 1421 & 32.638  & 347.5& 176.8$\pm$2.0 / 105.8$\pm$3.0 / ~56.0$\pm$3.5\\
1342239979  &  SPIREPhoto & 250/350/500 & 03-02 06:30:10 & 1421 & 32.627  & 306.8& 174.8$\pm$3.0 / 106.2$\pm$2.9 / ~57.7$\pm$3.5\\
1342240025  &  SPIREPhoto & 250/350/500 & 03-02 22:44:11 & 1421 & 32.617  & 268.8 &175.5$\pm$2.9 / 104.9$\pm$3.1 / ~58.6$\pm$3.5\\
1342241087  &  SPIREPhoto & 250/350/500 & 03-03 16:21:25 & 1421 & 32.606  & 227.4 &171.9$\pm$2.8 / 106.0$\pm$3.1 / ~59.4$\pm$3.6\\
1342241158  &  SPIREPhoto & 250/350/500 & 03-04 09:17:02 & 1421 & 32.596  & 187.6 &170.7$\pm$2.9 / 106.2$\pm$3.0 / ~56.5$\pm$3.5\\
1342240277  &  SPIREPhoto & 250/350/500 & 03-05 02:18:31 & 1421 & 32.585  & 147.6 &174.9$\pm$2.9 / 106.3$\pm$3.0 / ~56.5$\pm$3.5\\
1342240315  &  SPIREPhoto & 250/350/500 & 03-05 19:12:47 & 1421 & 32.574  & 107.9 &184.9$\pm$2.9 / 111.9$\pm$3.0 / ~59.8$\pm$3.5\\
1342240318  &  SPIREPhoto & 250/350/500 & 03-06 11:44:14 & 1421 & 32.563  &  69.1 &186.9$\pm$3.0 / 116.6$\pm$2.9 / ~63.9$\pm$3.5\\
1342241381-2 & PACSPhoto  & 70/160      & 03-14 03:02:01 & 2x286 & 32.442 & 358.7 & 321.9$\pm$8.6 /~~~~~~~~~~~~~~~~~~/ 331.0$\pm$10.1 \\
1342241383-4 & PACSPhoto  & 100/160     & 03-14 03:13:39 & 2x286 & 32.442 & 358.2 & ~~~~~~~~~~~~~~~~~/ 393.3$\pm$4.0 / 331.0$\pm$10.1\\
1342241418-9 & PACSPhoto  & 70/160      & 03-14 19:54:46 & 2x286 & 32.431 & 319.1 & 316.5$\pm$2.7 /~~~~~~~~~~~~~~~~~~/ 317.3$\pm$20.9\\
1342241420-1 & PACSPhoto  & 100/160     & 03-14 20:06:24 & 2x286 & 32.431 & 318.6 & ~~~~~~~~~~~~~~~~~/ 406.1$\pm$7.9 / 317.3$\pm$20.9\\
1342241471-2 & PACSPhoto  & 70/160      & 03-15 12:58:39 & 2x286 & 32.419 & 279.0 & 312.2$\pm$7.3 /~~~~~~~~~~~~~~~~~~/ 313.3$\pm$20.0 \\
1342241473-4 & PACSPhoto  & 100/160     & 03-15 13:10:17 & 2x286 & 32.419 & 278.6 & ~~~~~~~~~~~~~~~~~/380.4$\pm$13.4/ 313.3$\pm$20.0\\
1342241509-0 & PACSPhoto  & 70/160      & 03-16 06:33:59 & 2x286 & 32.407 & 237.6 & 295.0$\pm$7.1 /~~~~~~~~~~~~~~~~~~/ 314.8$\pm$13.6 \\
1342241511-2 & PACSPhoto  & 100/160     & 03-16 06:45:37 & 2x286 & 32.407 & 237.2 & ~~~~~~~~~~~~~~~~~/ 377.2$\pm$3.6 / 314.8$\pm$13.6\\
1342241620-1 & PACSPhoto  & 70/160      & 03-17 00:10:12 & 2x286 & 32.395 & 196.3 & 299.8$\pm$3.5 /~~~~~~~~~~~~~~~~~~/ 309.9$\pm$10.5\\
1342241622-3 & PACSPhoto  & 100/160     & 03-17 00:21:50 & 2x286 & 32.395 & 195.8 & ~~~~~~~~~~~~~~~~~/ 371.1$\pm$7.1 /  309.9$\pm$10.5\\
1342241655-6 & PACSPhoto  & 70/160      & 03-17 17:31:54 & 2x286 & 32.384 & 155.5 & 311.5$\pm$4.5 /~~~~~~~~~~~~~~~~~~/ 307.5$\pm$16.6\\
1342241657-8 & PACSPhoto  & 100/160     & 03-17 17:43:32 & 2x286 & 32.384 & 155.1 & ~~~~~~~~~~~~~~~~~/ 379.3$\pm$9.3 / 307.5$\pm$16.6\\
1342241699-0 & PACSPhoto  & 70/160      & 03-18 11:04:14 & 2x286 & 32.372 & 114.3 & 342.1$\pm$2.8 /~~~~~~~~~~~~~~~~~~/ 329.7$\pm$12.6 \\
1342241701-2 & PACSPhoto  & 100/160     & 03-18 11:15:52 & 2x286 & 32.372 & 113.9 & ~~~~~~~~~~~~~~~~~/ 418.3$\pm$2.9 /  329.7$\pm$12.6  \\
1342241865-6 & PACSPhoto  & 70/160      & 03-19 04:42:46 & 2x286 & 32.360 &  72.9 & 347.6$\pm$12.4 /~~~~~~~~~~~~~~~~/ 338.7$\pm$25.7\\
1342241867-8 & PACSPhoto  & 100/160     & 03-19 04:54:24 & 2x286 & 32.360 &  72.4 & ~~~~~~~~~~~~~~~~~/ 426.9$\pm$6.2 / 338.7$\pm$25.7\\
1342241928-9 & PACSPhoto  & 70/160      & 03-19 20:44:31 & 2x286 & 32.349 &  35.2 & 349.1$\pm$4.4 /~~~~~~~~~~~~~~~~~~/ 353.0$\pm$18.5\\
1342241930-1 & PACSPhoto  & 100/160     & 03-19 20:56:09 & 2x286 & 32.349 &  34.8 & ~~~~~~~~~~~~~~~~~/ 413.9$\pm$3.9 / 353.0$\pm$18.5\\
\hline
\multicolumn{8}{l}{$^a$ Observer-centric distance} \\
\multicolumn{8}{l}{$^b$ Subobserver east longitude at mid-point. We adopt the same orbital conventions as, e.g., \citet{buie97} and} \\
\multicolumn{8}{l}{~~\citet{lellouch11}. Zero longitude on Pluto is the sub-Charon point. The subobserver point longitude
decreases with time.}  \\
\multicolumn{8}{l}{$^c$ Color-corrected fluxes. PACS 160-$\mu$m fluxes are given for the average
over four consecutive Obs. IDs (see text)} \\
\end{tabular}
\end{center}
%\end{minipage}
\end{table*}

%__________________________________________________________________

\section{Herschel observations}
We obtained thermal photometry of the Pluto system with the two imaging photometers of 
{\em Herschel}, PACS (Photoconductor Array Camera and
Spectrometer; \citet{poglitsch10}) and SPIRE (Spectral and Photometric Imaging
Receiver; \citet{griffin10}), covering altogether six wavelengths.  The SPIRE instrument observes  a 4'x 8' field simultaneously in three bolometer arrays at 250 $\mu$m, 350 $\mu$m, and 500 $\mu$m, with respective pixel sizes of 6'', 10'', and 13''. PACS can operate in three filters, centered  at 70 $\mu$m (``blue"), 100 $\mu$m (``green"), and 160 $\mu$m (``red"). However, as it includes two detector arrays (64 x 32 pixels of 3.2''x 3.2'' for blue/green and 32 x 16 pixels
of 6.4'' x 6.4'' for red, each of them covering a FOV of 3.5' x 1.75'), only two filters out of three (70 / 160 $\mu$m or 100 / 160 $\mu$m) are observed in parallel.

For both instruments, the beam size (17"-35" FWHM for SPIRE and 5"-11" FWHM for PACS, depending on filter) encompassed the entire $\sim$1"-wide Pluto system, thus including thermal emission from Pluto and Charon (with a negligible contribution
from the other four moons). All data were acquired over three weeks in late February to mid-March 2012, under 
the \verb?OT2_elellouc_2? program (``Pluto's seasonal evolution and surface thermal properties").
%, 10.2 hours allocated in the {\em Herschel} Open Time 2 AO). 
We acquired nine observations of the Pluto system with each instrument. They
were timed to sample equally-spaced 
subobserver longitudes, so as to provide a multiband thermal light curve. In practice, consecutive 
visits to Pluto were scheduled with a time separation of $\sim$17 hours, equivalent to $\sim$40\deg\ longitude. The SPIRE observations occurred over Feb. 29 -- Mar. 6, 2012, while the PACS data were taken on Mar. 14--19, 2012. Pluto's heliocentric distance at that time was $r_h$ = 32.19 AU, the subsolar latitude was $\beta$ = 47.0\deg, and the phase angle was 1.6\deg.

We acquired the SPIRE observations  in the small-map mode. The telescope was
scanned across the sky at 30"/sec, in two nearly orthogonal (84.8\deg\ angle) scan paths, uniformly covering  an area of 5' x 5'. Each SPIRE visit to Pluto amounted to 1421 sec, corresponding to ten repetitions of the scanning pattern.
 
We used the mini scan map mode for PACS, which has been demonstrated to be more sensitive than the point-source (chop-nod) mode
\citep{muller10}\footnote{See also AOT Release Note: PACS Photometer Point/Compact Source Mode
2010, PICC-ME-TN-036, Version 2.0, custodian Th. M\"uller \citep{pacs10}.}. For each filter combination (70 / 160 $\mu$m or 100 / 160 $\mu$m), we acquired data consecutively in two scanning directions (termed ``A" and ``B"), with 70\deg\ and 110\deg\ angles with respect to the detector array and individual integration times of 286 sec per scan, i.e., 1144 sec (4 repetitions) per PACS visit. Observational details are given in Table~\ref{obssum}, where the A-B scanning sequences are indicated by consecutive Obs. ID numbers.

Far-infrared photometry can often be plagued by confusion noise, i.e., spatial variations in the sky emission at scales
comparable to the PSF. The confusion noise is typically $\sim$5-7 mJy/beam in the SPIRE bands \citep{nguyen10} and lower in the PACS bands, but Pluto's 2012 position in star-crowded regions of Sagittarius not far from Galactic center made sky background levels a priori more severe. Estimates of confusion levels at
proposal stage indicated that even though the March 2012 epoch was most favorable in this respect (and selected for
that reason), it would be subject to confusion noise at the $\sim$5 and $\sim$20 mJy in the PACS 100~$\mu$m and 160~$\mu$m beams,
respectively, i.e., a non-negligible fraction of the expected fluxes from Pluto ($\sim$400 and 300 mJy, respectively). However, the proper motion
of Pluto offered the possibility to observe the target several times against different sky backgrounds, permitting us to subtract  the sky contribution. The efficiency of this ``follow-on" (a.k.a. second-visit) approach has been demonstrated by the detection of numerous TNOs at the mJy level by {\em Spitzer} and {\em Herschel}  \citep[e.g.,][]{stansberry08, santos12}. For the technique to work, the proper motion between two visits should be significantly larger than the PSF size, but remain small enough that at the second visit, the object still falls within the high-coverage area of the map from the first visit. In practice, these conditions are best met for proper motions of 30''--50'' for PACS observations and 72"--150" for SPIRE.  In our observing sequence, the proper motion of Pluto between two consecutive visits ($\sim$17 hour separation) was of 55--35 arcsec, almost entirely in the RA direction (and decreasing with time as Pluto approached stationarity on April 10, 2012). Thus, for PACS observations, each visit to Pluto could be used as second epoch measurement for the preceding and/or following visit (17 hours before or after). For SPIRE, we often used more distant visits (i.e., 34 or 51 hours before or after a given observation) for the second epoch, as difference maps between two contiguous visits would result in the positive and negative Pluto images in the differential map to partially overlap at 500 $\mu$m.

\section{Data reduction}
{\it PACS}: Data reduction was initially performed within the {\em Herschel} Interactive Processing Environment \citep[HIPE;][]{ott10}, version 12, using its default FM7 calibration scheme and an optimum script for ``bright" sources.
% (I used the FM7 EEF-s to get the right fluxes).. 
Each PACS visit to Pluto provides two images (A and B scans) at 70 $\mu$m and 100 $\mu$m, and four images at 160 $\mu$m. 
For the green (100 $\mu$m) and red (160 $\mu$m) data, each image of a given visit was analyzed in combination with the corresponding image of the previous (``before") or successive (``after") visit to Pluto. The exception to this was, of course,  for the first (resp. last) visit to Pluto for which only the ``after" (resp. ``before") image could be used. This permitted us to generate two background maps, which were then subtracted from the individual image, providing a cleaner map suited for photometry. Standard aperture photometry on the resulting difference image was performed with our own IRAF/DAOPHOT-based routines via a curve-of-growth approach to determine the optimum synthetic aperture and a Monte-Carlo method of 200 fictitious source implantations to estimate error bars (see \citet{santos12} and \citet{kiss14} for details).
The method thus provided in general (i.e., except for the first and last visit, for which two times fewer values
were obtained) four (in the green) or eight (in the red) individual values of the flux (f$_i$, error bar $\sigma_i$) per visit.  The sky subtraction did not bring any noticeable improvement for blue (70-$\mu$m)
data, which have the least background contamination. Therefore, we simply performed aperture photometry on the original A and B images, providing two sets of values per visit. The optimum aperture radii were found to be 5.5", 7.0", and 10.5" in the blue, green, and red bands, respectively, i.e., close to the PSF
FWHM at the corresponding wavelengths. 
For each filter and visit, the 2 to 8 (1 to 4 for first and last visit) individually-determined fluxes were (error-bar weighted) averaged. To be conservative, we took
the final error on the average flux to be max (std(f$_i$), 1/$\sqrt{\sum \frac{1}{\sigma_i^2}}$), where std(f$_i$) is the standard deviation of the individual fluxes. Minor color corrections were finally applied, by dividing the averaged fluxes and their error bars by factors of 0.983 (70-$\mu$m), 0.982 (100-$\mu$m), and 1.000 (160-$\mu$m), appropriate (to within $\pm$0.01) for respective color temperatures of 47 K, 45 K, and 43 K. The final flux values are gathered in Table~\ref{obssum}. Additional systematic calibration uncertainties (5 \% of the measured flux), which do not affect the light curves, are not included in Table~\ref{obssum}.
 
% quadratically computed error which take into account the individual errors.

{\it SPIRE}: SPIRE data were first processed using HIPE, version 10, including de-striping routines that minimize background differences between data acquired at different epochs and that properly correct the signal timeline. Then, maps were produced using the standard naive map-making, projecting the data of each band on the same World Coordinate System (WCS) and applying cross-correlation routines between two epochs to correct for astrometry offsets.
Finally, for each Pluto visit, several difference maps were computed at each band by subtracting, from the map under consideration,
maps taken at other epochs, separated by $\sim$ $\pm$17, $\pm$34 and/or $\pm$51 hours (also depending on the considered filter). %fromthe considered map. 
Photometry on these difference maps was then performed 
% Contrarily to PACS, for SPIRE more distant visits, that is 34 hours before or after a given observations, had to be used as second epoch because the motion of Pluto was not big enough between two contiguous visits (i.e. separated by 17 hours), and the positive and negative Pluto fluxes in the differential map partially overlapped, especially at 500 $\mu$m.
%So for each visit and each band we have two background corrected maps (example for visit 4: map 4-2 and map 4-6) from which the Pluto flux was derived 
with a two-dimensional circular Gaussian aperture with a fixed filter-dependent FWHM (PSF fitting), following the method described in \citet{fornasier13}. The derived flux was then corrected by the instrument pixellization factors, i.e., dividing by 0.951, 0.931, and 0.902 for 250, 350, and 500 $\mu$m, respectively. Finally, color corrections were estimated by convolving   
%and color-corrected. For this step, we convolve 
a blackbody emission at 35--40 K, the approximate system brightness temperature at the SPIRE wavelengths, with the instrument 
spectral response profiles. These multiplicative color correction factors were found to be 0.974, 0.976, and 0.957 at 250, 350, and 500 $\mu$m,
and applied to the individual Pluto-Charon fluxes. The final system flux for each visit was then 
computed as the weighted mean of the individual fluxes based on the various differential maps (in a few cases after rejecting
some outliers). Final fluxes with all corrections included are gathered in Table 1. Similar to PACS, the errors include the uncertainties provided by the Gaussian fitting algorithm, but do not account for absolute calibration uncertainties, which are estimated to be 7\% of the measured flux.
The final PACS and SPIRE fluxes were converted into system brightness temperatures (T$_B$) by assuming a 1185 km radius for Pluto and 604 km for Charon. The Charon radius is based on stellar occultation \citep{sicardy06}. The adopted
Pluto radius is close to the best guess value from \citet{lellouch15}, 1184 km. These values match initial reports from {\em
New Horizons} \citep[606$\pm$3 km and 1187$\pm$4 km;][]{stern15}. Results would be insignificantly sensitive to further changes of the radii by a few kilometers.
The adopted value for Pluto's radius updates the value that was used in previous modeling of the ISO and {\em Spitzer} data 
(1170 km). The effect is negligible at 24 $\mu$m (a $\sim$0.1 K decrease in the T$_B$) but not entirely so at 500 $\mu$m ($\sim$0.5 K decrease).

%                                     Two column figure (place early!)

\begin{figure}[ht]
\centering
\includegraphics[width=9.cm,angle=0]{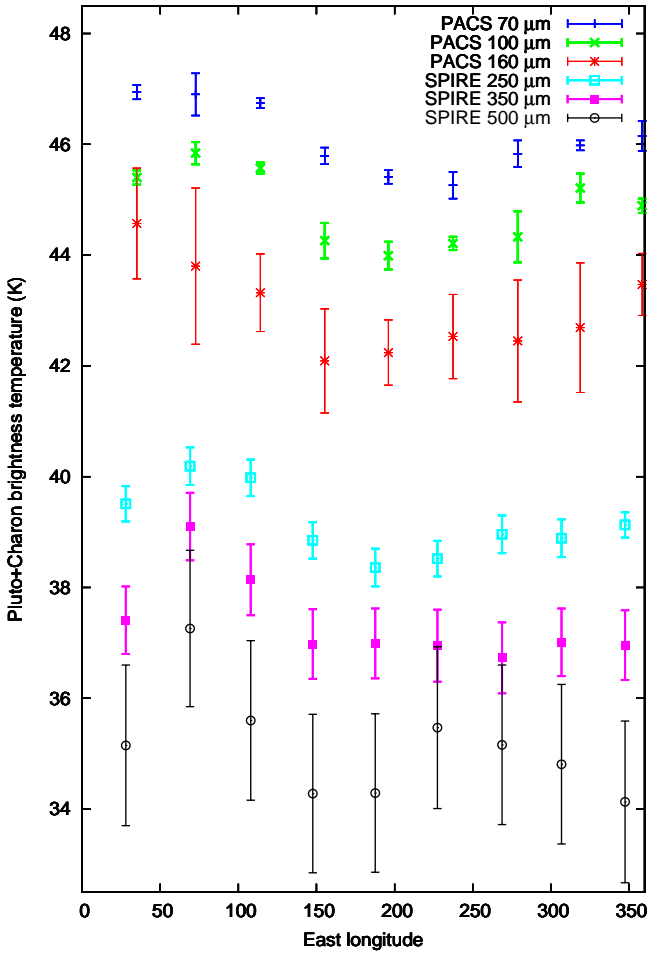}
\caption{Pluto-Charon thermal light curves in the six filters observed with PACS and SPIRE.}
\label{fig:herschel_data}
\end{figure}

%\vspace*{-1cm}

\section{Modeling}
\subsection{Qualitative analysis}
Brightness temperatures of the Pluto-Charon system over 70-500 $\mu$m as a function of rotational phase 
are shown in Fig.\ref{fig:herschel_data}. Immediately apparent in the figure is that: (i) the mean brightness
temperature (T$_B$) of the system decreases steadily with increasing wavelength from $\sim$46.5 K at 70 $\mu$m to $\sim$35 K
at 500 $\mu$m; and (ii) the thermal light curve is detected at all wavelengths, albeit somewhat marginally at 160 $\mu$m and especially
500 $\mu$m,
given the higher error bars of these data.
At 70 and 160 $\mu$m, the data are of much higher quality than was possible from {\em Spitzer} (see Fig. 2 from \citet{lellouch11}; hereafter Paper I). All data are consistent with maximum flux near an east longitude L = 60-80 and minimum flux near L = 200-220.
More quantitatitively, sinusoidal fits to the data yield flux maxima at L~=~57$\pm$5, 50$\pm$8, 42$\pm$22,  57$\pm$10, 
76$\pm$17, and 70$\pm$60 for 70, 100, 160, 250, 350, and 500-$\mu$m data. Thus, within measurements uncertainties,
all light curves appear in phase (in particular,  there is excellent phase agreement between the 70 $\mu$m, 100 $\mu$m and 250 $\mu$m
data). Out-of-phase light curves at the longest wavelengths had been envisaged in Paper I. 

\begin{figure*}[ht]
\centering
\includegraphics[width=13.cm,angle=-90]{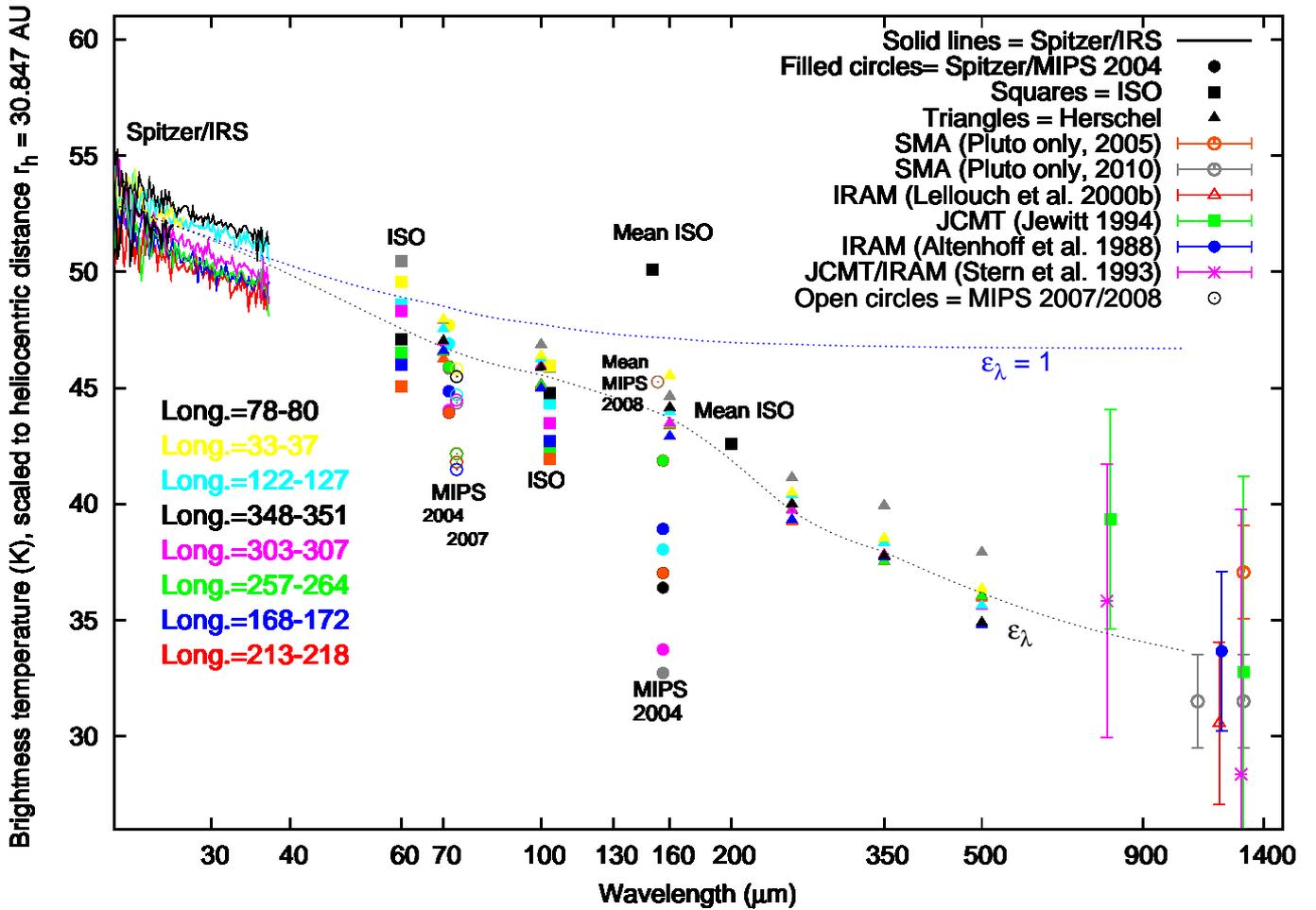}
\caption{Brightness temperature (T$_B)$ of the Pluto system. Most thermal observations from ISO, {\em Spitzer}, {\em Herschel}, and ground-based telescopes are gathered. As data were acquired at different epochs spanning $\sim$25 years,
the T$_B$ are rescaled by 1/$\sqrt{r_h}$ to a common epoch (September 2004, $r_h$ = 30.847 AU). 
Solid lines: eight {\em Spitzer}/IRS spectra over 21-37 $\mu$m, taken in Aug.-Sept. 2004 at east longitudes of 33 (yellow), 78 (gray), 
122 (light blue), 168 (dark blue), 213 (red), 257 (green), 302 (pink), and 348 (black). Filled circles: {\em Spitzer}/MIPS photometric measurements, taken at similar longitudes (37, 80, 127, 172, 218, 264, 307, and 351, same color codes) in September 2004. The {\em Spitzer} data are taken from \citet{lellouch11} (Paper I). Open circles: additional {\em Spitzer}/MIPS data at 71.42 $\mu$m from April 2007 (see Fig. 13
from Paper I).  Triangles: {\em Herschel} data at 70, 100, 160, 250, 250, and 500 $\mu$m from this work.  Filled squares: data from ISO at 60 and 100 $\mu$m taken in 1997 \citep{lellouch00a}. The {\em Herschel}, {\em Spitzer} 2007 and ISO data
are reinterpolated to the eight longitudes observed by {\em Spitzer} in 2004. The ISO 100 $\mu$m (resp. {\em Spitzer} 2007) data  are shifted by 3 $\mu$m (resp.  2 $\mu$m) for easier legibility. The comparison between the open and filled circles
at 71 $\mu$m illustrates the ``Pluto fading" witnessed by {\em Spitzer} from 2004 to 2007. Additional data (averaged over longitudes, no error bar) from ISO at 150 and 200 $\mu$m, and from unpublished {\em Spitzer} 156 $\mu$m observations taken in October 2008, are shown as filled squares and open circles respectively, under the labels ``Mean ISO" and ``Mean MIPS 2008".
For ground-based datasets sampling more than one longitude, only the average T$_B$ is plotted. The SMA-measured T$_B$
refers to Pluto only. The dotted lines show thermophysical model fits (see text), calculated for the conditions of September 2004.
Gray dotted line: parameters are from Case 4 in Table~\ref{emimodel}. Blue dotted line: same, but with spectral emissivities = 1. This latter case 
still produces brightness temperatures that decrease with wavelength, a consequence of the spatial mixing of different
surface temperatures.
}
\label{fig:overview}
\end{figure*}

In Fig. \ref{fig:overview}, the {\em Herschel} measured brightness temperatures are plotted as a function of wavelength and put into the broader context of most previous thermal measurements of the Pluto-Charon system. These measurements include (i) ISOPHOT 60, 100, 150, and
200 $\mu$m photometry, taken mostly in Feb.-March 1997 (five to eight visits to Pluto; \citet{lellouch00a}); (ii) {\em Spitzer}/MIPS 23.68, 71.42, and 156 $\mu$m photometry and {\em Spitzer}/IRS low-resolution spectroscopy over 20-37 $\mu$m recorded in August-September 2004 (eight visits to Pluto each; Paper I); (iii) additional {\em Spitzer}/MIPS data at 23.68 and 71.42 $\mu$m
taken in April 2007 (12 visits; see Fig. 13 of Paper I), and unpublished 156 $\mu$m data from October 2008
(12 visits); and (iv) a number of ground-based measurements at mm/sub-mm wavelengths from IRAM, JCMT, and SMA 
\citep{altenhoff88,stern93,jewitt94,lellouch00b,gurwell11}.  We emphasize that the SMA data separate Pluto from Charon, and we  report the Pluto-only T$_B$ from 2005 and 2010.  In Figure  
\ref{fig:overview}, {\em Spitzer}/MIPS and IRS data from 2004 at eight longitudes are plotted individually. We reinterpolate, to the same eight longitudes, the 
Herschel (70, 100, 160, 250, 350, and 500 $\mu$m), {\em Spitzer} 71.42 $\mu$m from 2007, and ISO 60 and 100 $\mu$m data, all of which clearly show light curves. For data in which we did not discern (or attempt
to detect) a light curve, i.e., ISO 150 and 250 $\mu$m, {\em Spitzer} 156 $\mu$m from October 2008, and all of the ground-based
data, we simply plotted the mean T$_B$ averaged over the available measurements. All of the ISO, {\em Spitzer,} and {\em Herschel} T$_B$ in Fig.~\ref{fig:overview} make consistent use of the above Pluto and Charon radii. In contrast,
mm/sub-mm T$_B$ simply use published values, because of the difficulty in tracking down the originally used radii. 
All these thermal measurements span  25 years
(1986-2012), a period over which Pluto's heliocentric distance ($r_h$) and subsolar latitude ($\beta$) varied from 29.7 AU to 32.2 AU and from -4\deg\ to +47\deg, respectively. While the effect of a change in the subsolar latitude cannot be accounted for
without a detailed model, the effect of varying $r_h$ is handled by rescaling the measured T$_B$ by 
1/$\sqrt{r_h}$ to the epoch of the {\em Spitzer} 2004 data ($r_h$ = 30.847 AU). 
 
Fig. \ref{fig:overview} illustrates
a number of important features. (i) The difficult-to-explain Pluto ``fading" witnessed by {\em Spitzer}, i.e., the decrease
by $\sim$2 K of the 71 $\mu$m T$_B$ (and by $\sim$0.5 K at 24 $\mu$m) over 2004-2007 (Paper I) is not confirmed
in the {\em Herschel} data, which indicates 70-$\mu$m T$_B$ in good agreement with the {\em Spitzer} 2004 data. (ii) The 150-160 $\mu$m T$_B$ show large dispersion. While the original ISO-150 $\mu$m data \citep{lellouch00a} indicated anomalously high fluxes 
(T$\geq$50 K in average),
the {\em Spitzer}/MIPS 156 $\mu$m data from April 2004 instead pointed to T$_B$ $<$ 40 K. The additional {\em Spitzer}/MIPS 156 $\mu$m unpublished data from October 2008 (12 visits) indicate a mean (rescaled) value of 45.3 K with a formal error of 1 K, but a 5.2 K dispersion over the 12 visits, which is a more likely representation of actual uncertainty. This mean value is generally in line,  albeit somewhat
on the higher side, with the 160 $\mu$m T$_B$ indicated by {\em Herschel.} (iii) The ensemble of data clearly outlines the decrease of the system brightness temperature with wavelength over the entire thermal range ($\lambda$ $>$ 20 $\mu$m). Although
data in the sub-mm/mm range show large dispersion, the most accurate of them (i.e., the SMA data from 2010 \citep{gurwell11}
 and the IRAM Feb.-Mar. 2000 data from \citet{lellouch00b}) point to a $\sim$32 K brightness temperature at 1100-1300 $\mu$m,
i.e., a consistent ``extrapolation"  of the trend indicated by the {\em Herschel} data  into the mm range. Thus it
appears that the Pluto-Charon T$_B$ decreases by more than 30 \% of its value from 20 $\mu$m ($\sim$53 K) to 500 $\mu$m ($\sim$35 K)
and beyond. 

\subsection{Emissivity}
Qualitatively, a decreasing T$_B$ with increasing wavelength
can be produced in several ways: (i) a spatially constant surface temperature T and a low (but spectrally constant) surface emissivity; (ii) the mixing of different surface temperatures. Such a mixing can occur both on regional
scales (e.g., different Pluto and Charon regions have different temperatures because
of different albedos or because they see different instantaneous insolations) and
on small scales (slopes at any scales, i.e., surface roughness, cause adjacent surface facets 
to see large variations of temperatures due to shadows and/or re-radiation); and (iii) a decrease of
the spectral emissivity with wavelength. Scenario (i) is technically possible, but fitting the
{\em Herschel}-measured brightness temperatures over 70-500 $\mu$m would require a
constant T$\sim$ 52.5 K and an improbably low ($\sim$0.57) spectrally constant surface emissivity. 
Rather, given the observed albedo variegation on Pluto and Charon, scenario (ii) must occur, at least 
on geographic scales. Based on multiterrain modeling of the {\em Spitzer} data, however, Paper I found  that
a decrease of the spectral emissivity with wavelength,  i.e. scenario (iii), of some of the Pluto areas (the CH$_4$ ice regions) was also
required.

Spectral ``emissivity" is often loosely defined in the literature, and sometimes treated as a fudge factor
in models. In essence, it represents the ratio of the observed fluxes to model fluxes, but how much physics is
put into the models leads to different estimates of the ``emissivity". In early works, surface temperatures
were described in simplistic end-member cases, such as the ``nonrotating" or the ``rapid rotator" cases. The advent of more
elaborate models, such as the asteroid STM \citep{lebofsky86} and NEATM \citep{harris98}, and of physically-based, thermophysical models \citep[TPM; e.g.,][]{spencer89, lagerros96, muller98}, provided a more realistic description of the surface temperature distribution across airless bodies, and thereby a definition of spectral emissivity in reference
to fluxes emitted from the {\em surface}. However, a further complication is that, particularly at long wavelengths, the surface 
materials' partial transparency may cause the emitted radiation not to originate at the surface itself, but from some characteristic
depth that depends on the material absorption coefficient. Therefore, the emitted flux does not just depend on the surface temperature, but on the thermal profiles $T(z)$ within the subsurface. Consideration of this aspect leads to another definition
of the emissivity, as the ratio of the observed to the modeled fluxes. The modeled flux, $\Phi_\nu$,
at some frequency $\nu$, is expressed locally as, e.g., \citep{keihm13} 
\begin{equation}
\Phi_\nu = \int~ B_\nu(T(z)) ~exp (-\frac{z}{L_e~ cos~\mu})~ \frac{dz}{L_e~ cos~\mu}
\end{equation}
and spatially integrated over the object. Here $B_\nu$ is the Planck function, $L_e$ is the electrical skin depth
(inverse of the absorption coefficient), and $\mu$ is the viewing geometry dependent angle between the outgoing radiation
and the surface normal. 

Many asteroid studies \citep[e.g.,][]{redman98,muller98}, making use of  {\em surface} temperature for reference,
reported evidence for very low spectral emissivities (0.6-0.7) in the mm/sub-mm ranges, which were attributed to grain 
size dependent subsurface scattering processes. However, the recent comprehensive study by \citet{keihm13} demonstrated that when allowance is made for subsurface sounding (and for the enhancement of the infrared fluxes by surface roughness), the observed mm/sub-mm fluxes are generally consistent with spectral emissivities close to 1 (e.g., 0.95).
This suggests that no scattering losses occur, except for specular reflection at the surface, 
which can be characterized by a Fresnel coefficient with moderate ($\sim$2.3) dielectric constant, characteristic of low-density material. As discussed below, however, there are other planetary surfaces where thermal scattering effects are demonstrated
to occur.

In our previous works \citep{lellouch00a,lellouch11}, the emissivity required to match the observed ISO or {\em Spitzer}
fluxes was defined in reference to a thermophysical model that only considered  the surface temperatures. 
A complication was related to the multiplicity of surface terrains. Three units (N$_2$ ice, CH$_4$ ice, tholin/H$_2$O ice) were considered for Pluto and one for Charon, and the approach was to (i) fix the spectral and bolometric emissivity of all units
except CH$_4$ and  (ii) adjust them for the CH$_4$ unit, using for initial guidance expectations based on spectral properties of ices
\citep{stansberry96}. Results (Paper I) suggested a large decrease of the spectral emissivity of CH$_4$
ice, from $\sim$1 at 24 $\mu$m to $\sim$0.4 at 200 $\mu$m. However, possible subsurface sounding effects were 
not considered, except in \citet{lellouch00b}, where the nondetection of the 1.2 mm light curve was interpreted
in terms of the mm-emissivity of the tholin/H$_2$O ice unit. 
Furthermore, all of our previous thermophysical models were ``diurnal-only", i.e., they assumed
equilibrium of the diurnally averaged temperatures with the instantaneous seasonal insolation. Thermal inertia
effects on seasonal timescales are thought to be important in controlling the surface-atmosphere exchanges
\citep{young12,young13,olkin15,hansen15} and the atmospheric pressure. They may be important to include
for our purposes because they are likely to impact the near-surface temperatures.
% in particular because the ices present on these bodies might be transparent enough for thermal radiation to probe below the diurnal skin depth.

Along with the decreasing brightness temperatures with increasing wavelength, 
a striking result of the {\em Herschel} measurement is the low T$_B$ value ($\sim$35 K) at 500 $\mu$m. Below we demonstrate
that such a T$_B$ is lower than any expected value for the dayside surface or subsurface of these bodies; thus, subsurface sounding toward the longest wavelengths is not the only cause of the decreasing T$_B$, so that  ``true" emissivity
effects (as defined per Equation (1)) must occur.

\subsection{Thermal modeling}
One complication associated with modeling of long-wavelength thermal data is related to the unknown level
probed by the emission within the surface, compared to the depth over which temperature changes occur 
(the thermal skin depth). Unlike in the thermal IR (e.g., 10-50 $\mu$m), where radiation is emitted from the surface
itself (within a fraction of a mm), materials can be transparent enough in the sub-mm that thermal
radiation might originate from layers $\sim$10 to several thousand times the wavelength (i.e., 5 mm to 1 m or more 
at 500 $\mu$m). The thermal skin depth is related to the thermal inertia $\Gamma$ through $d_s$~=~$\frac{\Gamma}{\rho~c}$$\sqrt{\frac{P}{\pi}}$, where $\rho$ is density, $c$ is heat capacity, and $P$ is the (diurnal or orbital) period. The parameter $d_s$ is therefore
not completely defined by $\Gamma$ since $\rho$ and $c$ may not be well known\footnote{It can be shown
(see, e.g., \citet{legall14} and \citet{schloerb15} for recent applications) that the outgoing thermal radiation depends on the ratio of the electric skin depth  $L_e$ to the relevant thermal skin depth, rather than on their absolute values.}. 
Still, using typical 
numbers for $\rho$ (900 kg m$^{-3}$) and $c$ (400 J kg$^{-1}$K $^{-1}$), a thermal inertia $\Gamma$~=~25 J~m$^{-2}$s$^{-0.5}$K$^{-1}$ (hereafter MKS) for Pluto leads to a diurnal skin depth of 3 cm, meaning that sub-mm radiation could 
either probe  within or below the diurnal skin depth and conceivably could even encompass a substantial fraction of the
seasonal skin depth (3.5 m for $\Gamma$ = 25 MKS). A much larger thermal inertia ($\Gamma$ = 1000-3000
MKS) could be more appropriate on seasonal timescales \citep{olkin15}, however; then
the seasonal skin depth would extend several hundreds of meters deep. 

Given these uncertainties,
our initial approach is to consider a number of situations of relative electrical, diurnal, and seasonal
skin depths. For that we first model horizontal and vertical temperatures of a ``mean" Pluto using a standard
spherical thermophysical model \citep{spencer89}\footnote{\url{https://www.boulder.swri.edu/~spencer/thermprojrs/}}. By ``mean Pluto", we mean that we use a Bond albedo of 0.46 derived from a mean geometric albedo $p_V$ = 0.58 and a  plausible
phase integral $q$~=~0.8 \citep{lellouch00a,brucker09}. A bolometric emissivity of 1.0 is assumed
and no surface roughness effects are included. As discussed below, the model is not relevant to the N$_2$-ice covered areas
whose heat budget is affected by sublimation/condensation terms.
 
\subsubsection{Diurnal-only models}
\begin{figure}[ht]
\centering
\includegraphics[width=6cm,angle=0]{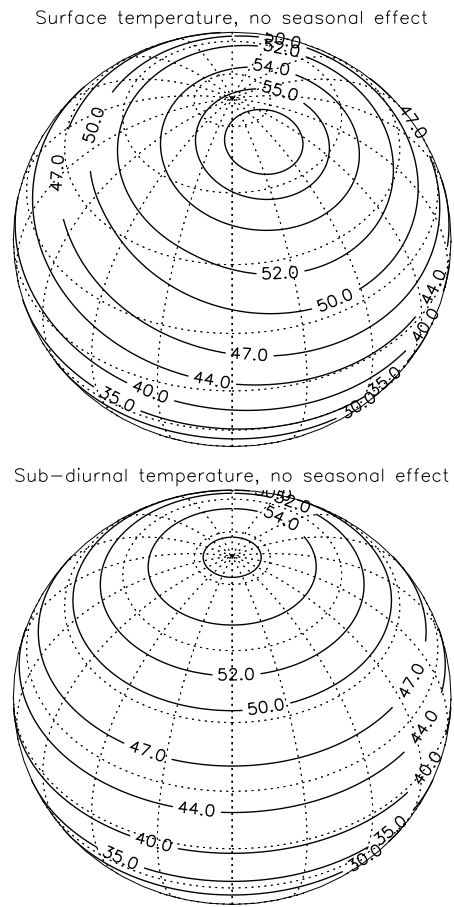}
\caption{Apparent Pluto temperatures, as viewed by a near-Sun observer in 2012, for a diurnal-only model with thermal inertia $\Gamma$ = 25 MKS, Bond albedo = 0.46
and bolometric emissivity $\epsilon_b$= 1. {\em Top:} Surface temperatures. {\em Bottom:} Temperatures at the bottom of the diurnal layer.}
\label{fig:diurnal_pluto}
\end{figure}

As a first step, we consider a ``diurnal-only" model. In this case, the local insolation is calculated using ``fixed"
heliocentric distance and subsolar latitude relevant to early March 2012, which leads in particular to zero temperatures
in the polar night southward of 43\deg S\footnote{This ignores internal heating. A typical radiogenic heating of 2.4 erg cm$^{-2}$s$^{-1}$ \citep{robuchon11}
would yield a $\sim$14 K polar night temperature for unit bolometric emissivity.}. A thermal inertia of 25 MKS is assumed, following results
from Paper I. Fig.~\ref{fig:diurnal_pluto} shows the resulting (i) surface temperatures and (ii)
``subdiurnal" temperatures, i.e., temperatures at depths much below the diurnal skin depth, in both cases as seen from the 
observer (neglecting the small 1.6\deg\ phase angle). Surface temperatures (relevant to dayside) peak near $\sim$56~K at high northern latitudes and fall below 35 K at latitudes below 20\deg~S only. Subdiurnal temperatures, which follow lines of equal
latitude, are slightly colder than the surface temperatures by 0-5 K (except in the 6-10 am morning hours where they can be warmer than surface temperatures by up to 4 K). Planck-averaged disk surface (resp. subdiurnal) temperatures over 70-500 $\mu$m are 51.4~--~49.0 K (resp. 49.8--47.0 K). All these temperatures are comfortably higher than the mean T$_B$ $\sim$ 35 K measured by {\em Herschel}-SPIRE at 500 $\mu$m, indicating that subsurface sounding within the diurnal layer is not the main culprit for this low T$_B$. 
Consideration of a possible positive thermal inertia gradient with depth in the diurnal layer would not change this conclusion because the subdiurnal temperature is an increasing function of thermal inertia \citep[e.g., Fig. 2 of][]{spencer89}.

\subsubsection{Seasonal models}
\begin{figure*}[ht]
\centering
\includegraphics[width=12cm,angle=90]{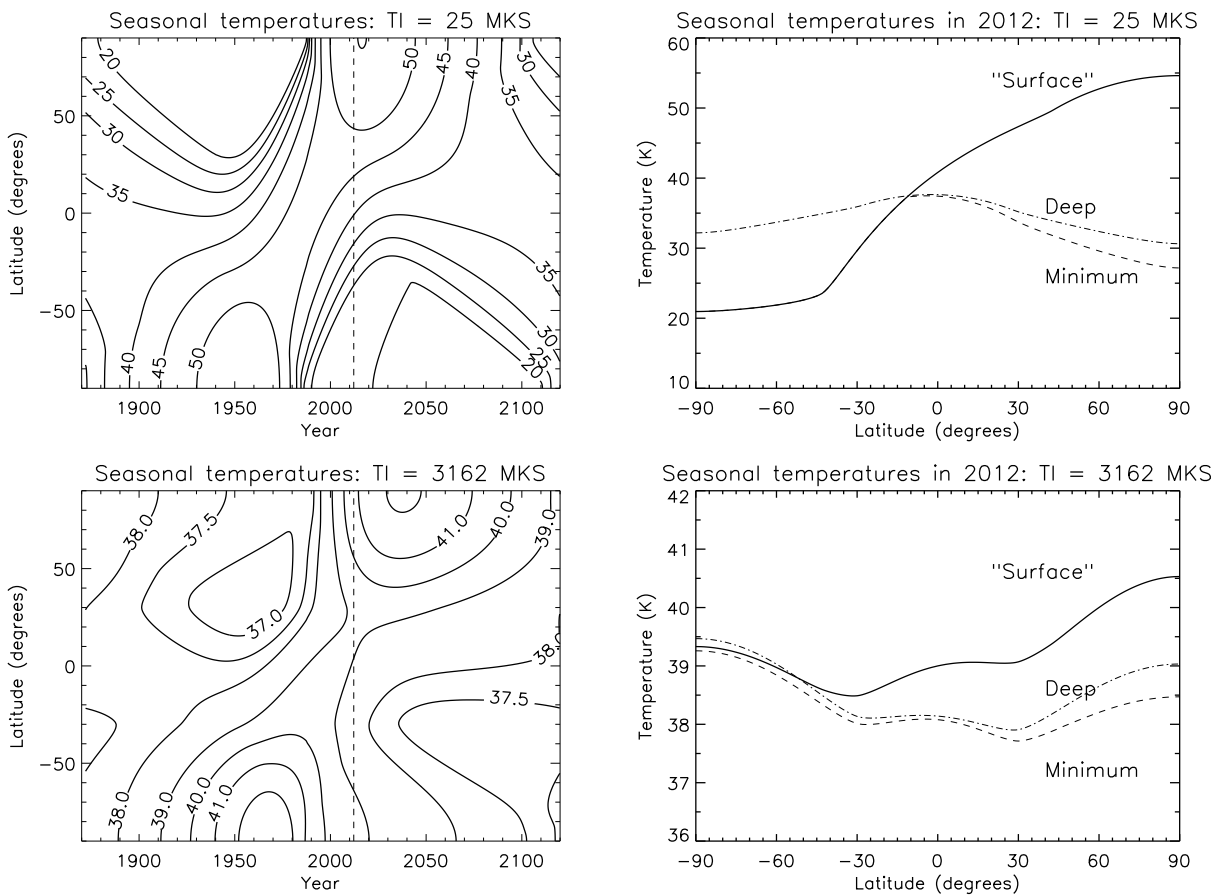}
\caption{Pluto temperatures from a seasonal model with thermal inertias $\Gamma$ = 25 MKS (top) and 3162 MKS (bottom). 
A Bond albedo of 0.46 and bolometric emissivity $\epsilon_b$= 1 are used. {\em Left:} ``Surface" temperature fields over a Pluto orbit. The dashed line indicates the epoch of the {\em Herschel} observations. {\em Right:} Temperatures as a function of latitude for early March 2012. ``Surface" and ``deep" indicate temperatures at the top and bottom of the seasonal layer. ``Minimum" refers to the minimum temperature within the seasonal layer for each latitude.}
\label{fig:seasonal_pluto}
\end{figure*}

The above approach does not consider the impact of thermal inertia on seasonal timescales. 
Continuing with a Bond albedo of 0.46 and bolometric emissivity  $\epsilon_b$= 1, we show in Fig.~\ref{fig:seasonal_pluto} the
seasonal temperature fields for two different values of the thermal inertia, $\Gamma$ = 25 MKS and 3162 MKS.
The first value, equal to that considered above for diurnal-only models, represents the situation of no thermal 
inertia gradient with depth. The second value represents one of the high thermal inertia cases favored by
some of the recent climate models \citep{young12,young13,olkin15}. In Fig.~\ref{fig:seasonal_pluto}, 
the left panels show the 2-D (time, season) diurnally averaged (i.e., subdiurnal) surface temperatures; the right panels, which
pertain to the epoch of the {\em Herschel} observations, show the latitudinal profile of: (i) this diurnally averaged temperature;
(ii) the ``deep" temperature (i.e., the temperature much below the seasonal skin depth); and (iii) the minimum value in the seasonal temperature vertical profile at each latitude. For the low thermal inertia ($\Gamma$ = 25 MKS) case, seasonal effects
on the diurnally averaged surface temperature are small. The temperature profile shown in the top right panel
of Fig.~\ref{fig:seasonal_pluto} closely matches the subdiurnal temperature map of Fig.~\ref{fig:diurnal_pluto}, 
except near and within the polar night where the zero temperatures of the diurnal-only
model are replaced by more physical $\sim$20 K-30 K values. In contrast, this model leads to rather cold temperatures
of 30-36 K in the ``deep" (i.e., subseasonal) subsurface, with minimum temperatures in the seasonal
layer occasionnally falling below 30 K at high northern latitudes.

The high ($\Gamma$ = 3162 MKS) thermal inertia case\footnote{The choice of 3162 MKS in \citet{young13} and \citet{olkin15} is just an effect of 
their thermal inertia grid, with two values per decade, for parameter
searches. No physical inference should be drawn from the fact that 3162 MKS is 
higher than the thermal inertia for solid H$_2$O at 40 K \citep[2200 MKS,][]{spencer92}.} 
has a much more dramatic effect on the diurnally averaged surface 
temperatures, which now show strongly subdued latitudinal contrasts of $\sim$3~K for south pole to north pole at the {\em Herschel} 
epoch. In this situation, the deep temperatures reflect the mean insolation over
the entire orbit and are almost hemispherically symmetric with maxima at the poles, minima near $\pm$30\deg latitude,
and a secondary maximum near the equator. These deep temperatures are in the range 38 K-39.5 K, and the minimum
vertical temperature never falls below 37 K.

\subsubsection{Comments and implications for the origin of low brightness temperatures}
The temperatures shown in Fig.~\ref{fig:diurnal_pluto} and \ref{fig:seasonal_pluto} are likely to be lower limits to the 
temperatures relevant to the {\em Herschel} observations for a number of reasons. First, they were calculated for
a bolometric emissivity of 1.0, which, if anything, minimizes the calculated temperatures. Second, the geometric albedo
that has been used is the Pluto-average value. Because the brightest regions are typically associated with N$_2$
ice, the non-N$_2$ ice regions are darker and thus warmer than the calculation indicates. Third, the above calculations
do not include any increase of the effective emitting temperature due to roughness \citep[those effects were incorporated in the form of 
a ``thermophysical model beaming factor" in][]{lellouch00a,lellouch11}. Finally, while the {\em Herschel} beam encompasses
Pluto and Charon, the above calculations pertain to Pluto only. Charon, which is slightly darker than Pluto, and
based on the {\em Spitzer} 24 $\mu$m data likely to have a slightly smaller thermal inertia in the diurnal layer (Paper I), should therefore be warmer than Pluto on its dayside. This argument cannot be applied to the seasonal models however, as the relative
seasonal thermal inertias of Pluto and Charon are unknown.

Yet, the temperatures shown in Fig.~\ref{fig:diurnal_pluto} and \ref{fig:seasonal_pluto} are only relevant to the Pluto units not covered by N$_2$ ice. The heat budget of the latter is dominated by sublimation-condensation exchanges, which, at a given point in time, maintain N$_2$ to an isothermal state over the globe and the surface pressure to an essentially constant value (except for 
topographic effects) that is buffered by the N$_2$ ice temperature \citep{young12}. The most recent volatile transport models 
\citep{young12,young13,olkin15,hansen15} %, constrained mostly by the pressure evolution over 1988-2013 
indicate N$_2$ ice temperatures constantly above 34 K throughout a Pluto year according to
\citet{hansen15}, and in the range 37.5-39.5 K according to \citet{olkin15}, with T(N$_2$) $\sim$
38.5 K in 2012. The surface pressure determination from New Horizons is $\sim$10~$\mu$bar in July 2015 \citep{stern15},
corresponding to equilibrium at 37.0 K \citep{fray09}. Because N$_2$ is horizontally isothermal (i.e., does not show
any diurnal temperature variation), it must also be vertically isothermal, at least over the diurnal skin depth. 
A firm lower limit of the N$_2$ temperature is provided by the shape of the (2-0) band at 2.15 $\mu$m,
which clearly indicates that N$_2$ is in the $\beta$ phase \citep{tryka94}, i.e., above the transition to cubic $\alpha$ phase
at 35.6 K \citep{scott76,trafton15}. All of this suggests that regions covered with N$_2$ ice are also warmer, albeit
not necessarily by much, than the mean 500-$\mu$m T$_B$ that we measure (35 K, Fig.\ref{fig:herschel_data}).

The above considerations show that in most situations physical temperatures at the surface and in the subsurface of Pluto
and Charon are considerably higher than the observed system brightness temperatures at 500 $\mu$m and beyond. The exception
is the case of the subseasonal temperatures for the case of low seasonal thermal inertia (25 MKS, i.e., comparable
to the diurnal seasonal thermal inertia), which can be as low as 27-36~K (Fig.~\ref{fig:seasonal_pluto}). 
We conclude that the low observed T$_B$
do not result from the temperature gradient (colder at depth on the dayside) in the diurnal layer, but could 
conceivably be due to long-wavelength radiation probing a significant portion of the seasonal layer. This 
situation would require that (i) the seasonal thermal inertia is small, i.e., there is no significant vertical gradient of the thermal inertia, and (ii) the surface material is transparent enough that thermal radiation probes several meters below the surface.

Estimates of Pluto's seasonal inertia have been obtained from climate models \citep{hansen96,young13,hansen15,olkin15} designed to match the atmospheric pressure evolution witnessed since 1988
and constraints on Pluto's albedo distribution based on HST measurements. In addition to thermal
inertia, these models include free parameters, such as the albedos and bolometric emissivities of the N$_2$ frost and the
involatile substrate, and the amout of volatile inventory. 
Once tuned to the pressure measurements, these models can also predict the orbit-long
evolution of Pluto's atmosphere. The latest two models, published prior to the {\em New Horizons} encounter, which
give different priorities on the constraints to fit, differ rather radically in their conclusions with contrasting best-fit solutions for the seasonal thermal inertia (10-42 MKS in \citet{hansen15}
vs 1000-3162 MKS for \citet{olkin15}) and diverging conclusions as to the fate of the atmosphere in the upcoming decades. The analysis of {\em New Horizons} data, particularly polar night temperatures with REX, should ultimately sort out this issue, but for now, we regard the seasonal thermal inertia of Pluto as significantly underconstrained. 
%we note that the Olkinet al. 2015 solution leads to ``deep" temperatures higher than we measure, requiring additional ``emissivity effects".

However, even if the seasonal thermal inertia is small (i.e., comparable to the thermal inertia in the diurnal layer), 
we believe that the sub-mm radiation does not probe a
large fraction of the seasonal skin depth (estimated above to be 3.5 m for $\Gamma$~=~25 MKS). This stems from our
estimate of the absorption coefficients of ices present on Pluto surface, on which we now elaborate. 
For N$_2$ ice and CH$_4$ ice,
\citet{lellouch00a} presented absorption coefficients over 30-300 $\mu$m based both on early laboratory data
compiled by \citet{stansberry96} and on new optical constants measurements. These measurements (see Fig. 8 of
\citet{lellouch00a}) indicate typical absorption coefficients of $\sim$0.5 cm$^{-1}$ for N$_2$ ice and
$\sim$1 cm$^{-1}$ for CH$_4$ ice, i.e., penetration depths of 2 cm and 1 cm, respectively, which is much shallower than
the above value of the seasonal skin depth. The significance of these penetration depths is actually uncertain because the volatile ices might actually be restricted to an even thinner surface veneer.

\begin{figure}[ht]
\centering
\includegraphics[width=6cm,angle=-90]{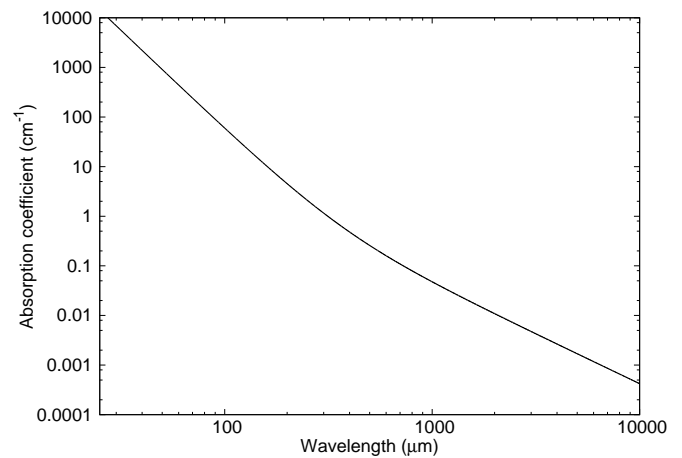}
\caption{Absorption coefficient of H$_2$O ice, extrapolated to 50 K (see text for details).}
\label{fig:coef_h2o_50K}
\end{figure}

H$_2$O ice on Pluto has long escaped spectroscopic detection, and based on initial {\em New Horizons} 
data appears to be exposed only in a number of specific locations, usually associated with red color, suggestive
of water ice/tholin mix \citep{grundy15,cook15}. Nonetheless, water ice is likely to be ubiquitous in Pluto's near
subsurface, given its cosmogonical abundance, Pluto's density, and its presence on Charon's surface\footnote{Evidence
for water ice bedrocks is also strongly suggested by the {\em New Horizons} discovery of several kilometer high topographic
features on both Pluto's and Charon's surfaces \citep{stern15}.}. 
Absorption coefficients for pure water ice ($k_{H_2O}$) at sub-mm-to-cm wavelengths are discussed extensively by \citet{matzler98},
who also provides several analytic formulations to estimate them as a function of frequency and
temperature along with illustrative plots. We use  the \citet{mishima83} formulation
\citep[see Appendix of][]{matzler98}. Its applicability is normally restricted to temperatures 
above 100 K, but Fig. 2 of \citet{matzler98}  indicates the trend with temperature. Absorption coefficients extrapolated to 50 K (estimated as half the values at 100 K) are shown in Fig.~\ref{fig:coef_h2o_50K}. 
%using the two formulations presented by M\"atzler (1998). 
At 500 $\mu$m, our best estimate is $k_{H_2O}$ = 0.25 cm$^{-1}$, comparable to the above values
for CH$_4$ and N$_2$ ices. The corresponding penetration length is therefore comparable to the diurnal skin depth but remains negligible compared to the seasonal skin depth, even for seasonal $\Gamma$ = 25 MKS. 
According to these calculations, the seasonal layer would be probed only at a 
wavelength of $\sim$4 mm and beyond. We also remark that the expression from \citet{mishima83}  would give a penetration depth
of 56 m at 2.2~cm, which is an order of magnitude larger than indicated by the laboratory measurements of \citet{paillou08}. In addition,
small concentrations of impurities can dramatically reduce the microwave transparency of water ice (e.g., \citet{chyba98} and references therein). Therefore, the above calculations likely indicate upper limits
to the actual penetration depth of radiation in a H$_2$O ice layer, from which we conclude that the seasonal layer is not
reached at the {\em Herschel} wavelengths.

We conclude that the low brightness temperatures observed at the longest {\em Herschel} wavelengths
cannot be explained by subsurface sounding, and imply emissivity effects. In what follows,
we present models aimed at fitting the {\em Herschel} light curves to evaluate
the mean spectral emissivity of the Pluto-Charon system.

\subsection{Fit of {\em Herschel} data}
Using the above thermophysical models, we expand upon the models developed previously for fitting the
ISO and {\em Spitzer} data \citep{lellouch00a,lellouch11}. Briefly, these models described the Pluto-Charon
system as composed of four units (N$_2$ ice, CH$_4$ ice, H$_2$O/tholin mix, and Charon), with specific
distributions and geometric and bolometric albedos constrained by Pluto's optical light curve, permitting
one to calculate the thermal radiation (assumed Lambertian) of the entire system. The distribution of surface units
was based on visible imaging and photometry (HST, mutual events) and near-infrared Earth-based spectroscopy.
To calculate the local surface temperatures,  a diurnal-only
thermophysical model was used for all four units except N$_2$ ice, which was maintained at a fixed
N$_2$ frost temperature. Another special condition was that the CH$_4$ temperature was allowed
to vary in accordance to thermophysical model predictions, but was capped at a maximum 54 K
temperature to account, in a simplified manner, for sublimation cooling effects for CH$_4$,
which become important above this temperature \citep{stansberry96b}. 

\begin{figure}[ht]
\centering
\hspace*{-0.5cm}\includegraphics[width=8.3cm,angle=0]{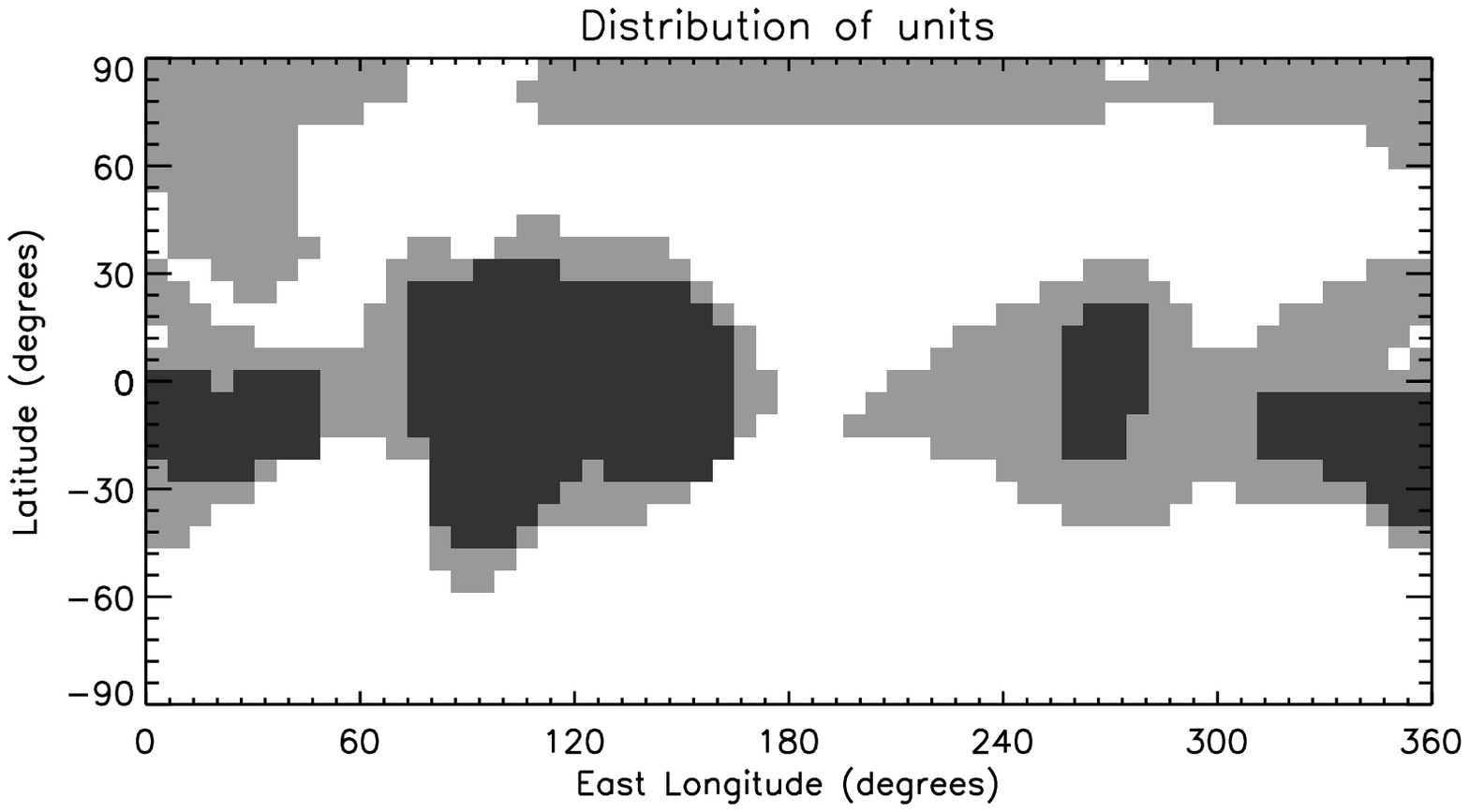}
%\hspace*{-0.5cm}\includegraphics[width=8.3cm,angle=0]{idl_s3.ps}
%\hspace*{.3cm}\includegraphics[width=9cm,angle=0]{distr_terrains_2015.ps}
\includegraphics[width=9cm,angle=0]{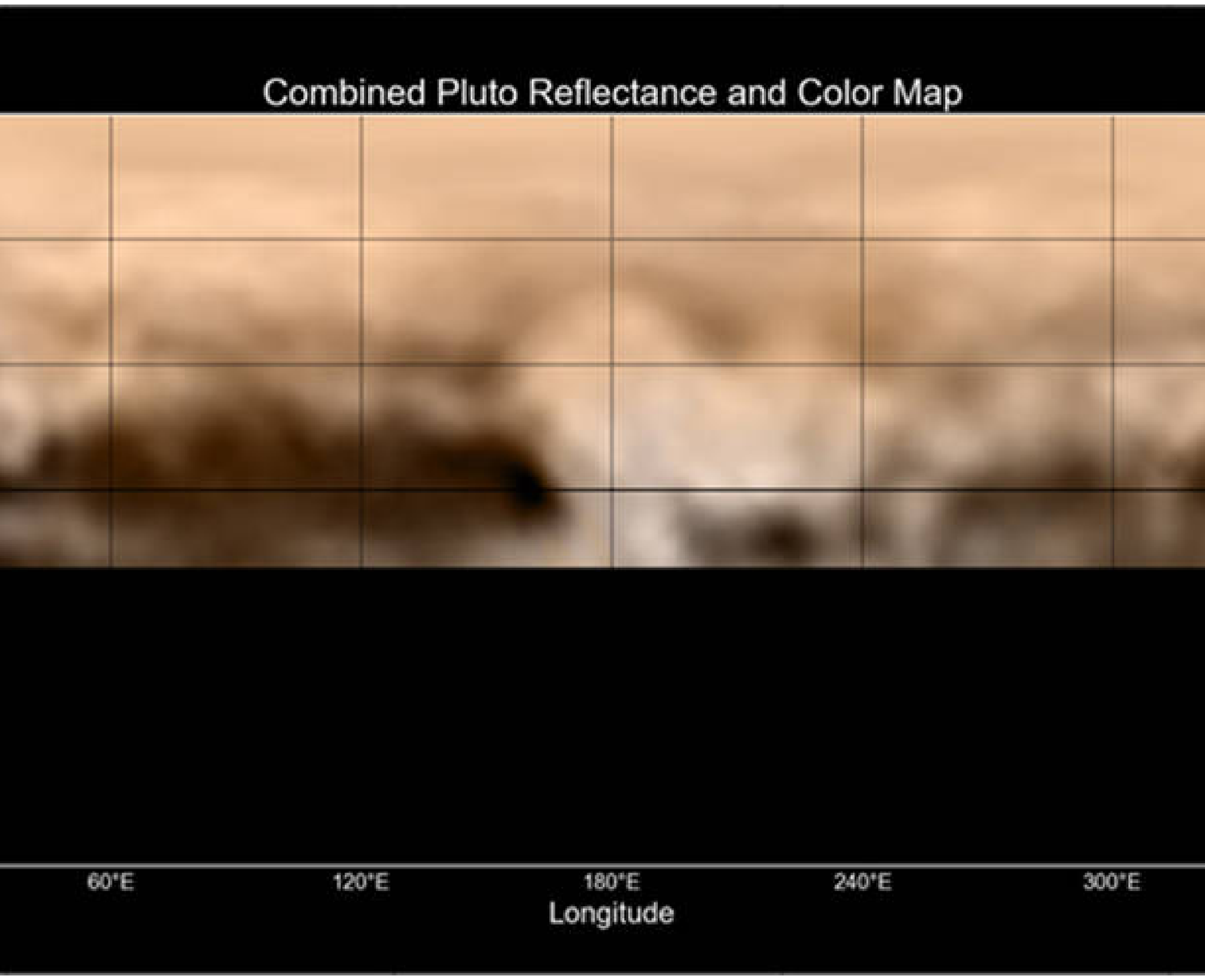}
\caption{Top: Adopted Pluto units for modeling (white = N$_2$, gray = CH$_4$, black = H$_2$O/tholin). Bottom:
Map of Pluto created from images taken from June 27 to July 3 by the Long Range Reconnaissance Imager (LORRI) on New Horizons, combined with lower resolution color data from the spacecraft's Ralph instrument. See http://pluto.jhuapl.edu/Multimedia/Science-Photos/pics/nh-pluto-map.jpg. Cthulhu Regio is the dark region covering $\sim$30$^{\circ}$E-160$^{\circ}$E longitudes. Sputnik
Planum is the southern part of the bright region immediately to the east \citep[informal names are taken from][]{stern15}.} 
\label{fig:units}
\end{figure}

This ``end-member" description is obviously outdated by the high-resolution {\em New Horizons}/LORRI imaging results \citep{stern15}, but until high-resolution maps of composition and albedo from LORRI and Ralph are available, it remains the only practical approach for our purpose. For now, we only considered the distribution favored in Paper I (``g$_2$'', their Fig. 4),  remarking its rather nice consistency with the early LORRI/images (Fig.~\ref{fig:units}). Furthermore, this distribution
is also roughly consistent with early compositional results from {\em New Horizons}/Ralph ,
which show both N$_2$ and CH$_4$ in Sputnik Planum, neither N$_2$ nor CH$_4$
in Cthulhu Regio, CH$_4$ north of Cthulhu and in the north polar region, and N$_2$ at mid-northern latitudes (\citet{grundy15}, L. Young, priv. comm). The model free parameters are the thermal inertias of Pluto and Charon (expressed in terms
of the thermal parameter $\Theta$ \footnote{$\Theta$ is related to the thermal inertia $\Gamma$ by $ \Theta=\frac{\Gamma\sqrt{\omega}}{\epsilon_b\sigma T_{SS}^3} $, where $\omega$ = 2$\pi$/(6.3872 days) is the body rotation rate, $\epsilon_b$ is the bolometric
emissivity of the surface, $\sigma$ is Stefan-Boltzmann's constant, and T$_{SS}$ is the instantaneous equilibrium
temperature at the subsolar point, }), plus the bolometric
and/or spectral emissivity of some of the units, especially CH$_4$ ice. Fitting the {\em Spitzer} 2004 light curve
makes it possible to estimate  the thermal inertias of 
Pluto and Charon separately because  the former primarily dictates the 24 $\mu$m light-curve amplitude, while the latter determines the 
large contribution of Charon to the observed mean 24-$\mu$m T$_B$ (see Paper I for details).

\begin{figure}[t]
\centering
\includegraphics[width=9cm,angle=0]{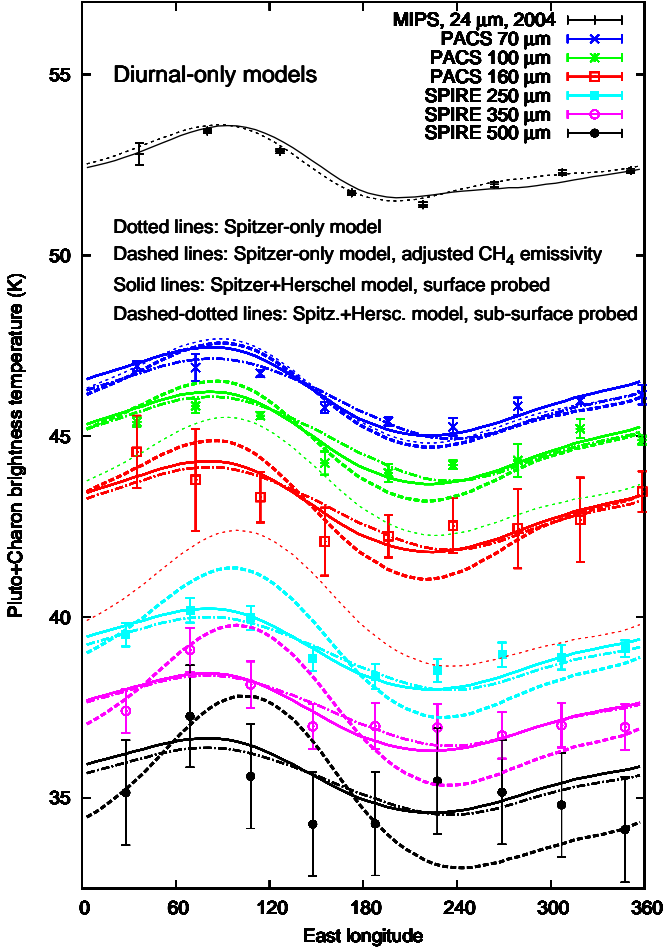}
\caption{Fits of the {\em Spitzer} 2004 (black points at top) and {\em Herschel} 2012 (all other points) brightness temperatures
of the Pluto-Charon system using a diurnal-only model. Thin dotted lines: {\em Spitzer}-derived model (case 1 in Table~\ref{emimodel}). Dashed lines:
{\em Spitzer}-derived model with CH$_4$ emissivities adjusted (case 2). Solid lines: New model (see text), assuming that
radiation originates at the surface at all {\em Herschel} wavelengths (case 4). Dashed lines: Same, but 
assuming that radiation originates in the subdiurnal layer at all {\em Herschel} wavelengths (case 5).}
\label{fig:fits_lc}
\end{figure}

We start by testing the best-fit model of Paper I determined from the {\em Spitzer} 2004 data. In this diurnal-only model, the spectral and bolometric emissivity of Charon and of the H$_2$O/tholin unit of Pluto were fixed to 1. Inferred parameters
were the thermal parameters of Pluto and Charon, $\Theta_{PL}$ = 6 (i.e., $\Gamma_{PL}$ = 22 MKS) and $\Theta_{CH}$ = 4.5
($\Gamma_{CH}$ = 22 MKS); bolometric emissivity of methane, $\epsilon_{b,CH_4}$ = 0.7; and spectral emissivity of methane,
$\epsilon_{CH_4}$ = 0.7, 0.6, and 0.45 at 70, 100, and 160 $\mu$m, respectively. As indicated by the 
thin dotted lines in Fig.~\ref{fig:fits_lc}, this model (case 1 in Table~\ref{emimodel}), which provides 
an excellent fit of the {\em Spitzer} 2004 data, is inconsistent with the {\em Herschel} measurements as it yields brightness temperatures that are too low  at 100 and 160 $\mu$m, as well as too much light-curve contrast at these wavelengths. The first deficiency
largely results from the poor quality of the {\em Spitzer} 2004 156 $\mu$m measurements. These measurements, which are now shown to be inconsistent with other data (Fig.\ref{fig:overview}), unduly skewed the model toward  brightness temperatures that are too low. 
This deficiency
can be corrected for by adjusting the CH$_4$ ice spectral emissivities (to 0.67, 0.80, 0.84, 0.58, 0.53, and 0.43
at 70, 100, 160, 250, 350, and 500 $\mu$m, respectively; case 2 in Table~\ref{emimodel}). 
However the synthetic light curves  (long dashed-lines in Fig.~\ref{fig:fits_lc}) still have too much contrast, except at 70 $\mu$m. This implies that the other units besides CH$_4$ ice are also subject to wavelength-dependent emissivities.
Relaxing the hypothesis that the spectral emissivity of Charon and of the H$_2$O/tholin units are equal to unity, we searched for the spectral emissivity (now assumed for simplicity to be the
same for Charon and the three Pluto units) that permits a fit to all light curves (case 3 
in Table~\ref{emimodel}). This case permits
a good fit (not shown in Fig.~\ref{fig:fits_lc}) to the data, but we do not regard it as satisfactory because
the associated  Planck-averaged bolometric emissivity is 0.82-0.85, which is inconsistent with the bolometric emissivities prescribed for 
the tholin/H$_2$O, CH$_4$ and Charon units (1.0, 0.7, and 1.0, respectively).

\begin{figure}[t]
\centering
\includegraphics[width=9cm,angle=0]{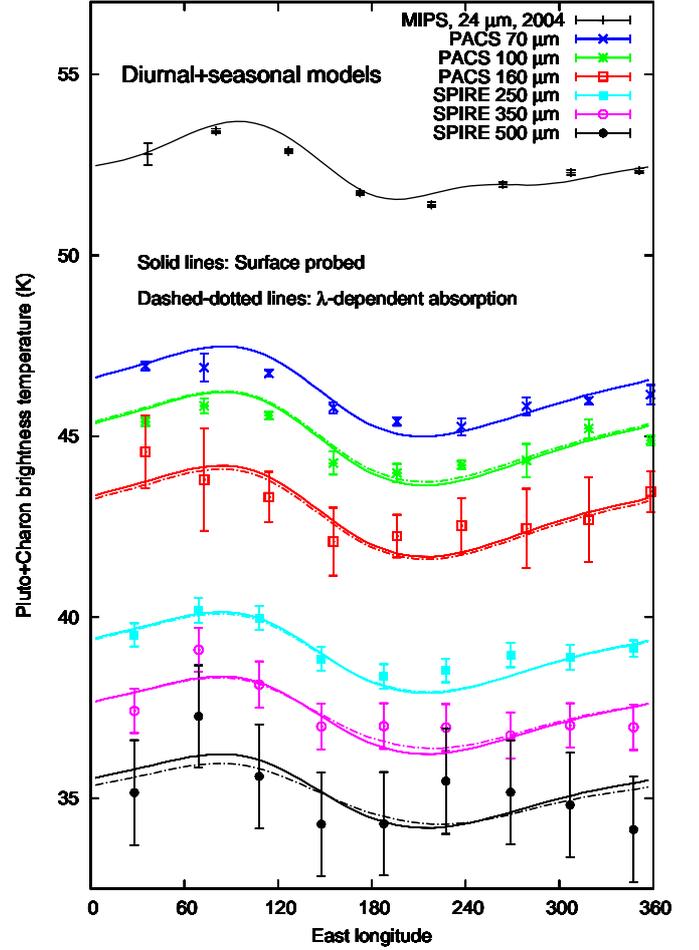}
\caption{Fits of the {\em Spitzer} 2004 (black points at top) and {\em Herschel} 2012 (all other points) brightness temperatures
of the Pluto-Charon system using a seasonal+diurnal model. The seasonal thermal inertia is 2000 MKS. Solid lines: Model assuming that
radiation originates at the surface at all {\em Herschel} wavelengths (case 4b in Table~\ref{emimodel}). Dashed lines: Same, but 
using the wavelength-dependent absorption coefficient from Fig.\ref{fig:coef_h2o_50K} (case 6b). 
}
\label{fig:fits_lc_seasonal}
\end{figure}

The above results point to the need to revise the {\em Spitzer}-derived models, and we here updated the fitting
approach. For the sake of simplicity, we adopted fiducial bolometric emissivities of 0.90 for Charon and all Pluto
units (for N$_2$ ice, the bolometric emissivity is not explicitly used; instead a uniform temperature is specified).
We also do not include a ``beaming factor" in the thermophysical model (as was done in Paper I), 
i.e., we ignore surface roughness effects; these are discussed separately later.
With these changes to the model, the {\em Spitzer}-2004 24 $\mu$m light curve was refit in terms of separate
thermal parameters for Pluto and Charon at the {\em Spitzer} epoch, adopting a 24 $\mu$m emissivity of 1.0 for all units. 
Best-fit $\Theta_{PL}$ = 7 and $\Gamma_{CH}$ = 3 values  (i.e., $\Gamma_{PL}$ = 26 MKS and $\Gamma_{CH}$ = 14 MKS) were obtained. The thermophysical model was then re-run
for the 2012 conditions, searching for the  spectral emissivities 
(again assumed to be the same for all units) that are permitted to fit
the ensemble of {\em Herschel} light curves. Because there is considerable uncertainty in the penetration
length of the far-IR radiation, we considered three cases: (i) small penetration at all six {\em Herschel}
wavelengths, i.e., the surface itself is probed (case 4); (ii) large penetration, i.e., the subdiurnal
layer is probed (case 5); and (iii) a wavelength-dependent absorption coefficient, following 
Fig.~\ref{fig:coef_h2o_50K} (case 6). The required spectral emissivities for cases 4-6 are 
given in Table~\ref{emimodel} and the associated emissivity curves are shown in Fig.~\ref{fig:emissivities}.
The overall fits of the thermal data for cases 4 and 6 are shown in Fig.~\ref{fig:fits_lc} with solid
and dashed-dotted lines, respectively.

Cases 4, 5, and 6 imply Planck-averaged bolometric emissivities of 0.83-0.86, 0.89-0.93, and 0.83-0.86. Although
this is not precisely consistent with the adopted bolometric emissivities of 0.90, and although these
models provide a somewhat worse fit to the 24 $\mu$m data than do the ``{\em Spitzer}-only" models, we consider
that the overall solution is physically satisfactory, and that Fig.~\ref{fig:emissivities} provides a proper
estimate of the spectral emissivity behavior of the Pluto-Charon system as a whole.

The above results pertain to diurnal-only models. To study the effect of a large seasonal thermal inertia on the
derived emissivities, cases 3 to 6 were reconsidered assuming that the subdiurnal temperature is determined
by a seasonal thermal inertia of 2000 MKS (for both Pluto and Charon). Solution cases  in terms
of the diurnal thermal inertias and spectral emissivities are given in Table~\ref{emimodel} (cases 3b to 6b). Since the large seasonal
thermal inertia implies cold subdiurnal temperatures, somewhat smaller thermal inertias (compared to the diurnal-only
case) are required to fit the {\em Spitzer} 24 $\mu$m fluxes. Furthermore, these subdiurnal temperatures are in this case
too cold (see also Fig.~\ref{fig:seasonal_pluto}) to fit the {\em Herschel} 70, 100, and 160 $\mu$m T$_B$, making this case (case 5b) not viable. In contrast, cases 3b, 4b, and 6b have emissivity solutions insignificantly different from corresponding cases 3, 4, 
and 6 (Table~\ref{emimodel}). Fits of the {\em Spitzer} 24 $\mu$m and {\em Herschel} data with these seasonal + diurnal
models (cases 4b and 6b) are shown in Fig.~\ref{fig:fits_lc_seasonal}. They are almost indistinguishable from those shown in 
Fig.~\ref{fig:fits_lc}. Thus (and not surprisingly given that all our data pertain to nearly the same season), we are unable to constrain the seasonal thermal inertia from our data. 
Nonetheless, the ensemble of model solutions (Table~\ref{emimodel}) yields diurnal thermal inertias $\Gamma_{PL}$ = 16-26 MKS and $\Gamma_{CH}$ = 9-14 MKS, confirming results from Paper I.

\begin{table*}
%\begin{minipage}[t]{\columnwidth}
\caption{Emissivity models} % title of Table
\label{emimodel}      % is used to refer this table in the text
\centering                          % used for centering table
\renewcommand{\footnoterule}{} 
\begin{center}
\begin{tabular}{|c|c|c|c|c|c|c|c|}
\hline
Case & $\Theta_{Pl}$$^a$ & $\Theta_{CH}$$^a$ & $\Gamma_{Pl}$ & $\Gamma_{CH}$ & $\epsilon_{b}$$^b$ & $\epsilon_{70,100,160,250,350,500\mu m}$$^c$ & Model type \\
\hline
1 & 6 & 4.5 & 22 & 22 & 1.0, 0.7, 1.0 & CH$_4$: .7,.6,.45, N/A, N/A, N/A$^d$            & surface temp. \\
2 & 6 & 4.5 & 22 & 22 & 1.0, 0.7, 1.0 & CH$_4$: .67,.80,.84,.58,.53,.43  & surface temp.\\
3 & 6 & 4.5 & 22 & 22 & 1.0, 0.7, 1.0 & all~~~: .84,.85,.83,.74,.72,.68  & surface temp.  \\
4 & 7 & 3 & 26 & 14   & 0.9, 0.9, 0.9 & all~~~:   .85,.86,.84,.74,.72,.70  & surface temp.  \\
5 & 7 & 3 & 26 & 14   & 0.9, 0.9, 0.9 & all~~~:   .95,.94,.89,.77,.75,.72  & sub-diurnal temp.  \\
6 & 7 & 3 & 26 & 14   & 0.9, 0.9, 0.9 & all~~:   .85,.86,.84,.75,.735,.71  & $\lambda$-dept. absorption \\
3b$^e$ & 4.5 & 2.5 & 16 & 12 & 1.0, 0.7, 1.0 & all: .84,.85,.825,.73,.715,.68  & surface temp.  \\
4b$^e$ & 4.5 & 2 & 16 & 9 & 0.9, 0.9, 0.9 & all~~~:   .87,.87,.84,.74,.72,.69  & surface temp.  \\
5b$^e$ & 4.5 & 2 & 16 & 9 & 0.9, 0.9, 0.9 & all~:   N/A,N/A,N/A,.91,.85,.80  & sub-diurnal temp.  \\
6b$^e$ & 4.5 & 2 & 16 & 9 & 0.9, 0.9, 0.9 & all~:   .87,.87,.84,.755,.74,.73  & $\lambda$-dept. absorption  \\

  \hline
\multicolumn{8}{l}{{\footnotesize $^a$ Thermal parameters at {\em Spitzer} 2004 epoch}}\\
\multicolumn{8}{l}{{\footnotesize $^b$ Bolometric emissivities of tholin/H$_2$O, CH$_4$ and Charon}} \\
\multicolumn{8}{l}{{\footnotesize $^c$ Spectral emissivities or either CH$_4$ only or all units}} \\
\multicolumn{8}{l}{{\footnotesize $^d$ {\em Spitzer} model: spectral emissivities of CH$_4$ only defined at 70, 100, and 160 $\mu$m}} \\
\multicolumn{8}{l}{{\footnotesize $^e$ Models with seasonal inertia $\Gamma$ = 2000 MKS. N/A in case 5b indicates no solution}}\\
 \end{tabular}

\end{center}
\end{table*}

Model predictions for Case 4 of Table~\ref{emimodel} over 20-1000 $\mu$m, calculated for the geometric conditions of September 2004 (subsolar latitude = 34.5\deg, heliocentric distance = 32.19 AU), are superimposed on Fig.~\ref{fig:overview} (gray dotted line). Although the figure gathers data with different subsolar latitudes, the fit shows the overall model adequacy. We do not attempt to fit the disparate set of sub-mm/mm data, noting that much improved constraints at these wavelengths are expected from ALMA \citep{butler15}. The same model, but in which spectral emissivities are forced to unity, is shown for comparison (blue dotted line). This latter case still shows a decrease of the brightness  temperatures with wavelength as a result of the mixing of different surface temperatures (see Section 4.2), but this effect is clearly not sufficient to explain the data.

In a recent study, Trafton (2015) questioned the paradigm interpretation of Pluto's 
near-infrared spectrum in terms of ``pure" and ``diluted" CH$_4$ ice, and, on thermodynamical grounds, 
proposed an alternate surface scenario with the mixture of areas covered by N$_2$-rich (N$_2$:CH$_4$, saturated with CH$_4$) 
and CH$_4$-rich (CH$_4$:N$_2$, saturated with N$_2$) solid solutions. For each unit, saturation of the secondary
component occurs at the several percent level \citep[about 3 \% at 37 K; see Table 1 of][]{trafton15}.
The CH$_4$:N$_2$ unit would correspond optically to what 
has been reported as ``pure CH$_4$", but with the key difference that this CH$_4$:N$_2$ unit would be isothermal because of the role of the N$_2$-rich ice in transporting latent heat between solid solutions. As suggested by 
Trafton, thermal measurements may provide a test of these ideas. Coming back to the diurnal-only
model (case 4 in Table~\ref{emimodel}) we attempted to remodel the {\em Spitzer} 24 $\mu$m light curve (best suited for this task because of its enhanced sensitivity to temperatures) under the assumption that the ``CH$_4$ ice unit" is actually isothermal at some constant temperature T$_{CH_4}$. As can be seen in Fig. 12 of \citet{lellouch11}, the contribution of the CH$_4$ unit is most important over L = 280--30 (and responsible for the increase of flux 
with increasing longitude in this range). If the other components (Charon and tholin/H$_2$O mix) 
are left untouched, the range of brightness temperatures measured in thislongitude bin requires T$_{CH_4}$ = 51-52 K, though the fit is not as good as it is with variable temperatures. 
Allowing for an increase in the contribution of the
tholin/H$_2$O or Charon unit (i.e., decreasing their thermal inertia) makes room for slightly
lower values of T$_{CH_4}$ ($\sim$ 50 K), but the shape of the calculated light curve
degrades unacceptably below this temperature. We conclude that the 24 $\mu$m light curve measured by Spitzer
(i) implies that if the ``CH$_4$ ice unit" actually represents isothermal CH$_4$-rich CH$_4$:N$_2$ mixtures,
these must be at least 50 K warm, i.e., much warmer than the  N$_2$-rich areas ($\sim$ 37 K); and (ii) favors spatially and diurnally variable temperatures for the CH$_4$-dominated areas over the isothermal case. At face value, these conclusions do not support
Trafton's (2015) scenario of isothermal CH$_4$:N$_2$ at the same temperature as N$_2$:CH$_4$, although reconciliation might be possible if regions attributed to pure CH$_4$ actually represent a spotty coverage of CH$_4$:N$_2$ 
solutions at 37 K with nonvolatile material. Finally, at 50 K, the saturated N$_2$ abundance in
CH$_4$:N$_2$ would be 2--3 times larger than at 37 K.

\section{Discussion}
\subsection{Roughness effects?}
\begin{figure}[ht]
\centering
\includegraphics[width=6cm,angle=-90]{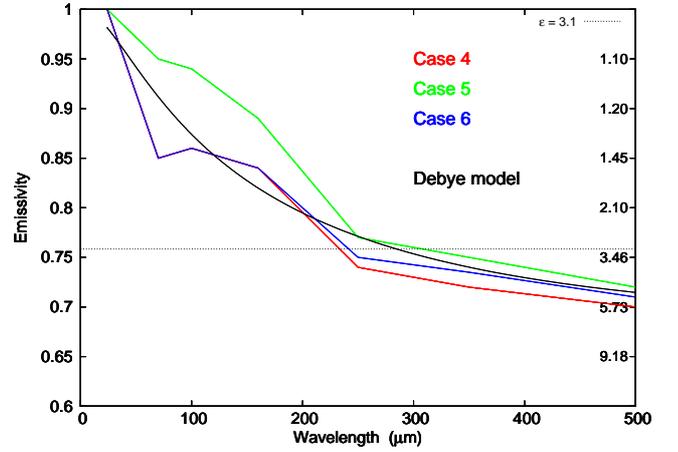}
\caption{Derived emissivities for models 4-6 (see text and Table~\ref{emimodel}). These emissivities
are compared to predictions for a smooth surface with dielectric constants indicated on the right y scale (the dotted line
is for a dielectric constant of 3.10) and for
a Debye model with $\epsilon$$_s$ = 5, $\epsilon_\infty$ = 1 and $\nu_r$ = 10~$cm^{-1}$ (see text). }
\label{fig:emissivities}
\end{figure}

The emissivities derived in this work (Table~\ref{emimodel}, Fig. \ref{fig:emissivities}) either make  use of
a simple treatment of surface roughness in the case of the ``{\em Spitzer}-only" models (cases 2 and 3 in Table~\ref{emimodel})
or ignore roughness (cases 4, 5, 6, 4b, 6b). 
As reviewed, e.g., in \citet{keihm13} and \citet{delbo15}, disk-integrated infrared measurements of airless bodies have long indicated
flux enhancements in near zero-phase angle observations relative to thermophysical models of smooth
surfaces (even if zero thermal inertia is used in these models). These flux enhancements, strongest
at shorter thermal wavelengths, are commonly viewed as the effect of small-scale surface roughness. 
The latter results in a multiplicity of surface temperatures at any scale with an enhanced
contribution of the hottest temperatures to the flux, particularly at shorter wavelengths. These effects led
to the introduction of the ``beaming factor" in the literature \citep[e.g.,][]{lebofsky86,harris98,delbo15},
whereby a semiempirical correction to the thermophysically calculated surface temperatures can be applied
to match the infrared fluxes. While this approach is usually appropriate for fitting disk-averaged observations in terms
of an object's diameter and albedo, it becomes insufficient when dealing with multiwavelength,
multiphase angle, and/or multi local-time data. Examples of its shortcomings have been demonstrated, for example, 
by comparing its predictions of center-to-limb temperature profiles to those of physical roughness models
constrained by lunar thermal emission profiles \citep{rozitis11}. Other evidence
of thermal emission enhancements not amenable to a single ``beaming factor" was obtained from 
spectral images of comets 103P/Hartley 2 and 9P/Tempel 1 \citep{groussin13} at thermal wavelengths. These
data indicate color temperatures that are barely dependent on incidence angle $i$, vastly exceeding
predictions from smooth surface models at large values of $i$. This is interpreted by the fact that
thermal emission from a given region of a comet is dominated by facets  
that are oriented toward the Sun with a temperature that is mostly independent of the ``smooth" 
incidence angle, but instead strongly dependent on local topography (slopes, projected shadows) on sub-km
to sub-mm scales. In \citet{groussin13}, these effects were modeled by replacing the Planck 
function B($\lambda$, $T$) by a product $\Lambda$$\times$B($\lambda$, $T$),
where $T$ is the color temperature and $\Lambda$ ($<$1) represents in essence the fraction of an
observed region that undergoes the highest temperatures,  as measured by $T$. The $\Lambda$ parameter was found to 
decrease with increasing incidence angle, as expected for progressively larger effects of projected
shadows. 
%As outlined by Groussin et al. (2013), this is not sufficient to handle the complexity of unresolved multiple surface temperatures within a resolution element of the model. In particular, this approach leads to a global increase of the local temperature, and cannot account for the fact that at any observational scale, the color and brightness temperatures are different, the latter decreasing with increasing wavelength.

In many asteroid thermal models, macroscopic roughness (e.g., occurring on scales larger than the thermal skin depth) is usually
described by crater models, accounting for effects of partial shadowing, scattering of sunlight, mutual
radiative heat exchanges within depressions, using different approximations and methods of coupling with heat conduction.
Key parameters are the crater surface coverage and depth-to-diameter ratio, which combine to define
the surface ``rms slope". For example, \citet{lagerros98} found that high roughnesses (rms slope $\sim$35\deg)
are required to mimic, in a flux-averaged sense, the standard beaming factor for asteroids. Based on the crater formulation from \citet{hansen77}, \citet{keihm13}'s calculations for a typical 0.15 albedo asteroid at 2.5 AU with low thermal inertia indicate
that large roughnesses (50 \% coverage of hemispherical craters) produce flux enhancements (over the smooth surface case) by
$\sim$9 \% at 100 $\mu$m, but as much as $\sim$40 \% at 12 $\mu$m (see their Fig. 1)\footnote{Similar numbers are obtained
from the crater model of \citet{muller02}.} and even more at shorter wavelengths. These calculated flux enhancements can be applied to other objects by noting that they are unique functions of $\lambda$$\times$T by virtue of the Planck function dependence. ``Transposing" a 0.15 albedo asteroid at r$_h$~=~2.5 AU (which has an instantaneous
subsolar temperature of T$_{SS}$ = 246.0 K for $\epsilon$=0.9) to Pluto (A~=~0.46, r$_h$~=~32.2 AU, giving T$_{SS}$ = 61.2 K),
means that for equal roughness the same flux enhancements would occur for Pluto over 48~$\mu$m-400 $\mu$m. Although
the case described by \citet{keihm13} presumably represents an upper limit of any realistic roughness for Pluto, 
the above comparison might suggest that wavelength-dependent roughness effects may affect the {\em Spitzer}+{\em Herschel} fluxes at typical levels of a few tens of percent. While the enhancements due to roughness are wavelength-dependent
when considered in flux, they in fact imply approximately constant increases in brightness temperatures.
Specifically, when reference is made to a smooth equilibrium model (EQM), the above flux enhancements for the considered
asteroid imply brightness temperatures increases of 14-15~K over 12-100 $\mu$m. Rescaling to the Pluto
case would mean that the 48~$\mu$m-400~$\mu$m T$_B$ could be affected by roughness effects at the 3-4 K level at most
with essentially no spectral dependence\footnote{For comparison, the ``thermophysical beaming factor" used in Paper I was 0.905-0.925, corresponding to a typical $\sim$1 K increase of the brightness temperatures.}. This is a small fraction of the observed T$_B$s decrease ($\sim$15 K) over that
interval (Fig. \ref{fig:overview}). The potential 3-4 K effect is even dwarfed by the 10~K T$_B$ difference associated 
with a spectral emissivity of $\sim$0.7, as we find required by the 500 $\mu$m T$_B$ (Table~\ref{emimodel}). Finally and 
most importantly, and as alluded to in Section 4.3.3, since surface roughness can only enhance disk-averaged fluxes,
any significant effects would actually exacerbate (by a few K) the fact, outlined in Section 4.3, that the long-wavelength
T$_B$ are below any plausible temperatures within Pluto's subsurface. We are left to conclude that roughness effects
are not the cause of the emissivity spectral dependence that we observe.

\subsection{Comparison to other bodies and interpretation}
Modeling of the {\em Herschel} data has led us to (i) an updated estimate of the Pluto and Charon 
diurnal thermal inertias, and (ii) an assessement of their spectral emissivities over 20-500 $\mu$m.
We determine $\Gamma_{PL}$ = 16-26 MKS and $\Gamma_{CH}$ = 9-14 MKS, in good agreement from inferences
based on {\em Spitzer}-only data (Paper I), namely $\Gamma_{PL}$ = 20-30 MKS and $\Gamma_{CH}$ = 10 -150 MKS (with
most solutions calling for $\Gamma_{CH}$ = 10-20 MKS). These thermal inertias are low compared to those of compact ices,
but still factors-of-several higher than the value statistically determined for the TNO population \citep[$\Gamma$ = 2.5$\pm$0.5 MKS,][]{lellouch13b}. As discussed in that paper, the difference does not necessarily imply intrinsically different thermal surface properties. For equal density and conduction properties, the diurnal skin depth at Pluto/Charon (P = 6.387 day) is $\sim$5 times
larger than that of a typical TNO with an 8-hour rotation period; hence the difference between Charon and a typical
TNO might simply be consistent with an approximately linear increase of the thermal inertia with depth. The apparently
higher thermal inertia at Pluto vs Charon may be related to atmospheric-assisted conduction in a porous surface \citep{lellouch00a}.

The very large ($>$30 \%) decline of the Pluto-Charon brightness temperatures from $\sim$20 to 500 $\mu$m, and probably beyond, is partly caused by the mixing of different temperatures on regional scales. The main conclusion of our work is
that, once this is taken into account, the reminder of the effect is not caused by surface roughness or subsurface
sounding at the longest wavelengths, so that ``genuine" emissivity spectral dependence occurs. This conclusion
confirms and expands that reached for the 1.2 mm emission of the Pluto-Charon system \citep{lellouch00b}.
 
Low emissivities at long wavelengths have been observed on a number of solar system icy bodies. \citet{muhleman91}
reported 3-40 mm flux measurements of Europa and especially Ganymede indicating brightness temperatures well below
the subdiurnal temperature, confirming earlier results from \citet{depater84}. Using Cassini/RADAR, \citet{ostro06} 
determined 2.2 cm brightness temperatures of several Saturn satellites. When comparing these brightness temperatures to the ``isothermal equilibrium 
temperature" (i.e., the mean equilibrium surface temperature over a sphere), this implied averaged
emissivities as low as 0.44 for Tethys, 0.59 for Enceladus and Rhea, and 0.69-0.81 for Iapetus. A 0.6-0.7
emissivity was also found at Enceladus by \citet{ries15}. However,  these estimates did not
include the effect of subsurface sounding. Focusing on Iapetus, but using a thermophysical modeling including subsurface
sounding, \citet{legall14} determined slightly higher emissivities (0.78-0.87, depending on the regions), but still
significantly below unity. Also at Iapetus, \citet{ries12} determined an extraordinarily low 9 mm effective
emissivity ($\sim$0.3-0.4) on the trailing side. Emissivity effects are also seen in millimeter-wavelength 
measurements of various kinds of ice and snow on Earth \citep{hewison99}. 

Fewer emissivity measurements are available in the far-IR (as opposed to sub-mm/cm wavelengths). From ISO/LWS observations of Mars, \citet{burgdorf00} inferred a spectral emissivity declining from 0.97 at 50 $\mu$m to 0.92 at 180 $\mu$m. Using {\em Herschel}, \citet{leyrat12} inferred
a large decrease of the spectral emissivity of asteroid 4 Vesta, from 0.9 at 70 $\mu$m to 0.7 at 500 $\mu$m, essentially
confirming previous findings by \citet{muller98}. These analyses, however, while including a detailed surface temperature model, did not account for possible subsurface sounding effects. Evidence for a spectrally-decreasing emissivity was also found for several Kuiper Belt and Centaurs by \citet{fornasier13}  from combined {\em Spitzer}/MIPS, {\em Herschel}/PACS, and {\em Herschel}/SPIRE data. Although once again, this work did not explicitly include vertical temperature profiles, a striking observational fact was the abrupt fall-off of the emitted fluxes beyond 300 or 400 $\mu$m, with most objects not detected at 500 $\mu$m.

Our results for Pluto-Charon add further evidence that unlike dust/rock regolith asteroids \citep{keihm13}, icy solar 
system surfaces show long-wavelength emissivity effects not amenable to a combination of surface roughness and subsurface 
sounding. Reasons for lower-than-unity emissivities may include (i) dielectric constants larger than 1, implying
reflection of the upward thermal radiation at the surface interface; (ii) particle scattering, which
produces an emissivity minimum for particle sizes comparable to $\lambda$/4$\pi$; and  (iii) volume scattering, whereby
the combination of a weakly-absorbing medium down to the electrical skin depth and inhomogeneities or voids on scales comparable or larger than the wavelength causes internal reflections \citep{ostro06,legall14}. 
Volume scattering is commonly invoked as the dominant scattering mechanism at microwave (mm-cm) wavelengths \citep{janssen09, legall14, ries15}. 

In the measurements of \citet{hewison99}, the ice/snow emissivity dependence with frequency varies with the age, wetness,
surface state, and transparency of the ice as a result of the combination of dielectric and scattering effects. This
was modeled in a semiempirical way using a Debye-like form of the complex permittivity,

\begin{displaymath}
\epsilon(\nu) = \frac{\epsilon_s - \epsilon_\infty}{1 - ~i~\nu / \nu_r} + \epsilon_\infty
\end{displaymath}
from which Fresnel coefficients can be calculated as a function of incidence angle. 
Here $\epsilon$$_s$ is the effective static permittivity, $\epsilon_\infty$ its high-frequency
limit and $\nu_r$ is the effective relaxation frequency.  This parameterization
handles both dielectric surfaces (by setting $\epsilon$$_s$ $>$ $\epsilon_\infty$)
and volume scattering ($\epsilon$$_s$ $<$ $\epsilon_\infty$), and both behaviors are
found in terrestrial icy material. Adopting the above parameterization, and assuming a smooth surface
and an equal mix of the two polarizations when calculating the Fresnel coefficients, the spectral dependence we derive for Pluto-Charon's emissivity can be approximately fit with $\epsilon$$_s$ = 5, $\epsilon_\infty$ = 1, and $\nu_r$ = 10 cm$^{-1}$ 
(Fig.~\ref{fig:emissivities}). The values of $\epsilon$$_s$ and $\epsilon_\infty$  encompass that of the water ice dielectric constant \citep[3.10-3.13 at 50-100 K;][]{gough72,paillou08}, for which a $\sim$0.76 constant spectral emissivity would be expected. Nonetheless, the above should be seen primarily as a working empirical model, and  $\epsilon$$_s$ $>$ $\epsilon_\infty$ suggests that volume scattering may not be important in causing the depressed emissivities
over 70-500 $\mu$m. In fact, as the volume scattering process operates in the electrical depth layer and for 
voids/inhomogeneities larger than the wavelength, it cannot be important when the absorption coefficient
becomes smaller than the inverse of the wavelength, i.e., below 100 $\mu$m for H$_2$O ice.

The emissivity decrease with wavelength, possibly extending toward the mm range, may also indicate 
particle scattering with a typical particle size of at least 100 $\mu$m. \citet{stansberry96} performed emissivity 
calculations for N$_2$ ice and CH$_4$ ice with various grain sizes, based on Hapke theory \citep{hapke93} and their far-IR absorption properties. Their results do indicate significantly less than unity far-IR emissivities, but the spectral behavior, with an emissivity decrease occurring only longward of 50 $\mu$m for CH$_4$ and 150 $\mu$m for N$_2$, is not consistent with the mean
emissivity behavior we infer here. A similar problem was noted in Paper I, where the high 24 $\mu$m emissivity
inferred for CH$_4$ ice was inconsistent with the calculations of \citet{stansberry96} for a broad range of 
grain sizes. Calculations for H$_2$O ice are not available but the similarity of its absorption coefficient
to that of N$_2$ and CH$_4$ ice at $\sim$300 $\mu$m (about 1 cm$^{-1}$) indicates that IR emissivities lower than unity 
are to be expected. Notwithstanding with the above issues, we conclude that the mean emissivity curve we have derived likely
results from the combination of a high dielectric constant and particle scattering in relatively transparent surface ices. 

%The temperature of Pluto's N$_2$ ice can be inferred from the atmospheric pressure: the buffering temperature of a 10 $\mu$bar N$_2$ atmosphere \citep{stern15} is 37.0 K \citep{fray09}. Assuming a bolometric emissivity of 0.90
%for N$_2$, the Bond albedo that leads to a globally averaged temperature (equal to T$_{SS}$/$\sqrt{2}$) of 37.0 K at
%32.9 AU (July 2015) is 0.70. This is not far from the geometric albedo (0.73) derived in Paper I for the N$_2$ unit on the basis
%of the assumed terrain distribution (``g2"), given their adopted phase integral (0.9). A bolometric emissivity of 0.90 is also
%reasonably consistent with the preferred climate models from \cite{olkin15} (models PNV9 and PNV12 from \cite{young13},
%which have a 0.80 emissivity. In contrast, the prefered model from \cite{hansen15} has a frost
%emissivity of only 0.6. Photometric (albedos, phase integrals) investigations from {\em New Horizons} will
%shed further light on this problem.

The temperature of Pluto's N$_2$ ice can be inferred from the atmospheric pressure. The buffering temperature of a 10 $\mu$bar N$_2$ atmosphere \citep{stern15} is 37.0 K \citep{fray09}. An expression for the globally constant temperature of the
N$_2$ ice (T$_{N_2}$) can be found in \citet{stansberry99} as a function of $\gamma$, the ratio of the total area of N$_2$ ice to
its cross-sectional area as viewed from the Sun. For ubiquitous N$_2$ frosts (and many other reasonable
surface ice configurations), $\gamma$ = 4, and T$_{N_2}$ =  T$_{SS}$/$\sqrt{2}$. For r$_h$ = 32.9 AU (July 2015), T$_{N_2}$ = 37.0~K implies a relationship between the Bond albedo (A$_b$) and bolometric emissivity, namely (1-A$_b$)~/~$\epsilon_b$ = 0.335. For the N$_2$ ice unit of the assumed terrain distribution (``g2"), Paper I derived a geometric albedo of 0.73, and assumed a phase integral of 0.90, yielding A$_b$ = 0.657. With a bolometric emissivity $\epsilon_b$ = 0.9 as we advocate here, 
this yields (1-A$_b$)~/~$\epsilon_b$ = 0.381. The agreement with the above value is only approximate but could be brought to perfection with very small changes, e.g., using $\gamma$ = 4.5 instead of 4 or a phase integral of 0.96 instead of 0.90. 
Furthermore, our nominal value of 0.381 matches well the range of solutions derived from the climate models \citep{young13,hansen15}. Indeed, restricting the discussion to models that best approach a $\sim$10 $\mu$bar pressure in 2015 (cases PNV21-23 and EPP14 from \citet{young13}, and models \#55 and \#66 from \citet{hansen15}), these models all converge to (1 - A$_b$) / $\epsilon_b$ in the range 0.363--0.375. Thus, it appears that there is no major difficulty with a N$_2$ bolometric emissivity of 0.9.
The heat budget for N$_2$ ice will be best revisited after results from photometric (albedos, phase integrals) investigations from {\em New Horizons} become available.

Although the {\em New Horizons} spacecraft does not carry a  dedicated  thermal radiometer, 
the Radio Experiment (REX) acquired measurements of the thermal emission from Pluto at 4.2 cm during two linear scans across the disk at close range including both day and night sides, and a third scan was obtained during the dark side transit of the occultation \citep{linscott15}. These data should provide crucial information on the surface temperature 
and its spatial variations, especially on the polar night temperature, which is the most diagnostic of thermal inertia on seasonal timescales (see Fig. 4). The present work suggests that emissivity effects will strongly impact the interpretation
of these data.

The fits of the {\em Herschel} data presented here should still be viewed as preliminary. %They should be superseded by 
Future analyses making use of detailed surface maps based on {\em New Horizons}/Ralph (including albedo, composition and its
vertical statigraphy, particle size, phase functions, and possibly temperatures from band shape) will be possible when those datasets 
become available \citep[e.g.,][]{grundy15}. More generally, the ISO, {\em Spitzer,} and {\em Herschel} combined Pluto-Charon light curves constitute a legacy dataset at thermal wavelengths, against which temperature predictions from seasonal climate
models should be tested, which will be soon complemented by additional data. Rotationally
resolved ALMA data separating Pluto from Charon are already available \citep{butler15}, and thermal maps at $\sim$0.01"
spatial resolution will be achievable in the coming years. Starting in 2018, the JWST/MIRI will measure
Pluto and Charon emission over $\sim$10-27 $\mu$m. These facilities will be invaluable to study the predicted
evolution of the surface thermal properties as the Pluto system recedes from the Sun.

\begin{acknowledgements}
PACS has been developed by a consortium of institutes led by MPE (Germany) and including UVIE (Austria); KU Leuven, CSL, IMEC (Belgium); CEA, LAM (France); MPIA (Germany); INAF-IFSI/OAA/OAP/OAT, LENS, SISSA (Italy); IAC (Spain). This development has been supported by the funding agencies BMVIT (Austria), ESA-PRODEX (Belgium), CEA/CNES (France), DLR (Germany), ASI/INAF (Italy), and CICYT/MCYT (Spain).
SPIRE has been developed by a consortium of institutes led by Cardiff University (UK) and including Univ. Lethbridge (Canada); NAOC (China); CEA, LAM (France); IFSI, Univ. Padua (Italy); IAC (Spain); Stockholm Observatory (Sweden); Imperial College London, RAL, UCL-MSSL, UKATC, Univ. Sussex (UK); and Caltech, JPL, NHSC, Univ. Colorado (USA). This development has been supported by national funding agencies: CSA (Canada); NAOC (China); CEA, CNES, CNRS (France); ASI (Italy); MCINN (Spain); SNSB (Sweden); STFC, UKSA (UK); and NASA (USA).
Additional funding support for some instrument activities has been
provided by ESA.
Data presented in this paper were analyzed using ``HIPE'', a joint
development by the {\em Herschel} Science Ground Segment Consortium,
consisting of ESA, the NASA {\em Herschel} Science Center, and the HIFI,
PACS, and SPIRE consortia. 
We are indebted to the ESA staff, in particular Rosario Lorente and
Mark Kidger, for a flexible and flawless implementation of the observations. E.L., S.F., and R.M. were supported by the
French Programme National de Plan\'etologie. C.K. and G.M. were supported by the OTKA/NKFIH K-104607 grant of
the Hungarian Research Fund / National Research, Development and
Innovation Office; and the contract 4000109997/13/NL/KML of the
Hungarian Space Office and the European Space Agency. Support for the work of
S.P. was provided by NASA through an award issue by JPL/Caltech.
We thank Leslie Young for constructive comments. 

\end{acknowledgements}

%----------------------------------

\end{document}